\documentclass[prl,twocolumn,showpacs,preprintnumbers,
superscriptaddress,amsmath,amssymb]{revtex4-1}
\usepackage{amsfonts, amsmath, bm, bbm, graphicx, epsfig, epstopdf}
\usepackage{anyfontsize}
\usepackage{dcolumn}
\usepackage{textcomp}
\usepackage{titlesec}
\usepackage{float}	
\usepackage[rflt]{floatflt}
\usepackage{leftidx}
\usepackage{siunitx}
\usepackage{esvect}
\renewcommand{\figurename}{\footnotesize{\textbf{Figure}}}

\titleformat{\section}[runin]
  {\normalfont\large\bfseries}{\thesection}{1em}{}
\titleformat{\subsection}[runin]
  {\normalfont\large\bfseries}{\thesubsection}{1em}{}
\begin{document}

\bibliographystyle{apsrev}

\title{Possible existence of optical communication channels in the brain}

\author{Sourabh~Kumar}
\affiliation{Institute for Quantum Science and Technology and Department of Physics
and Astronomy, University of Calgary, Calgary T2N 1N4, Alberta, Canada}
\author{Kristine~Boone}
\affiliation{Institute for Quantum Science and Technology and Department of Physics
and Astronomy, University of Calgary, Calgary T2N 1N4, Alberta, Canada}
\author{Jack~Tuszy{\'n}ski}
\affiliation{Department of Oncology, University of Alberta, Cross Cancer Institute, Edmonton T6G 1Z2, Alberta, Canada}\affiliation{Department of Physics, University of Alberta, Edmonton  T6G 2E1,
Alberta, Canada}
\author{Paul~Barclay}
\affiliation{Institute for Quantum Science and Technology and Department of Physics
and Astronomy, University of Calgary, Calgary T2N 1N4, Alberta, Canada}
\affiliation{National Institute for Nanotechnology, Edmonton T6G 2M9, Alberta, Canada}
\author{Christoph~Simon}
\affiliation{Institute for Quantum Science and Technology and Department of Physics
and Astronomy, University of Calgary, Calgary T2N 1N4, Alberta, Canada}
\date{\today}
\begin{abstract}
Given that many fundamental questions in neuroscience are still open, it seems pertinent to explore whether the brain might use other physical modalities than the ones that have been discovered so far. In particular it is well established that neurons can emit photons, which prompts the question whether these biophotons could serve as signals between neurons, in addition to the well-known electro-chemical signals. For such communication to be targeted, the photons would need to travel in waveguides. Here we show, based on detailed theoretical modeling, that myelinated axons could serve as photonic waveguides, taking into account realistic optical imperfections. We propose experiments, both \textit{in vivo} and \textit{in vitro}, to test our hypothesis. We discuss the implications of our results, including the question whether photons could mediate long-range quantum entanglement in the brain.
\end{abstract}

\maketitle

The human brain is a dynamic physical system of unparalleled complexity. While neuroscience has made great strides, many fundamental questions are still unanswered \cite{bigquestions}, including the processes underlying memory formation \cite{memory}, the working principle of anesthesia \cite{anesthesia}, and--most fundamentally--the generation of conscious experience \cite{koch,tononi,tegmarkCSF}. It therefore seems pertinent to explore whether the brain might generate, transmit and store information using other physical modalities than the ones that have been discovered so far.

In the present work we focus on the question whether biophotons could serve as a supplementary information carrier in the brain in addition to the well established electro-chemical signals. Biophotons are the quanta of light spanning the near-UV to near-IR frequency range. They are produced mostly by electronically excited molecular species in a variety of oxidative metabolic processes \cite{popp,cifra} in cells. They may play a role in cell to cell communication \cite{popp,fels}, and have been observed in many organisms, including humans, and in different parts of the body, including the brain \cite{isojima,kobayashi,tang1,kataoka}. Photons in the brain could serve as ideal candidates for information transfer. They travel tens of millions of times faster than a typical electrical neural signal and are not prone to thermal noise at body temperature owing to their relatively high energies. It is conceivable that evolution might have found a way to utilize these precious high-energy resources for information transfer, even if they were just the by--products of metabolism to begin with. Most of the required molecular machinery seems to exist in living cells such as neurons \cite{grass}. Mitochondrial respiration \cite{zhuravlev,tuszynski1} or lipid oxidation \cite{mazhul} could serve as sources, and centrosomes \cite{buehler} or chromophores in the mitochondria \cite{kato} could serve as  detectors.

However, one crucial element for optical communication is not well established, namely the existence of physical links to connect all of these spatially separated agents in a selective way. The only viable way to achieve targeted optical communication in the dense and (seemingly) disordered brain environment is for the photons to travel in waveguides. Mitochondria, and microtubules in neurons have been hypothesized to serve as waveguides \cite{thar,rahnama,jibu1,jibu2}. However mitochondria are typically less than a few microns long, and microtubules are much too thin to guide light in the biophotonic wavelength range.

\begin{figure}
\includegraphics[scale=0.08]{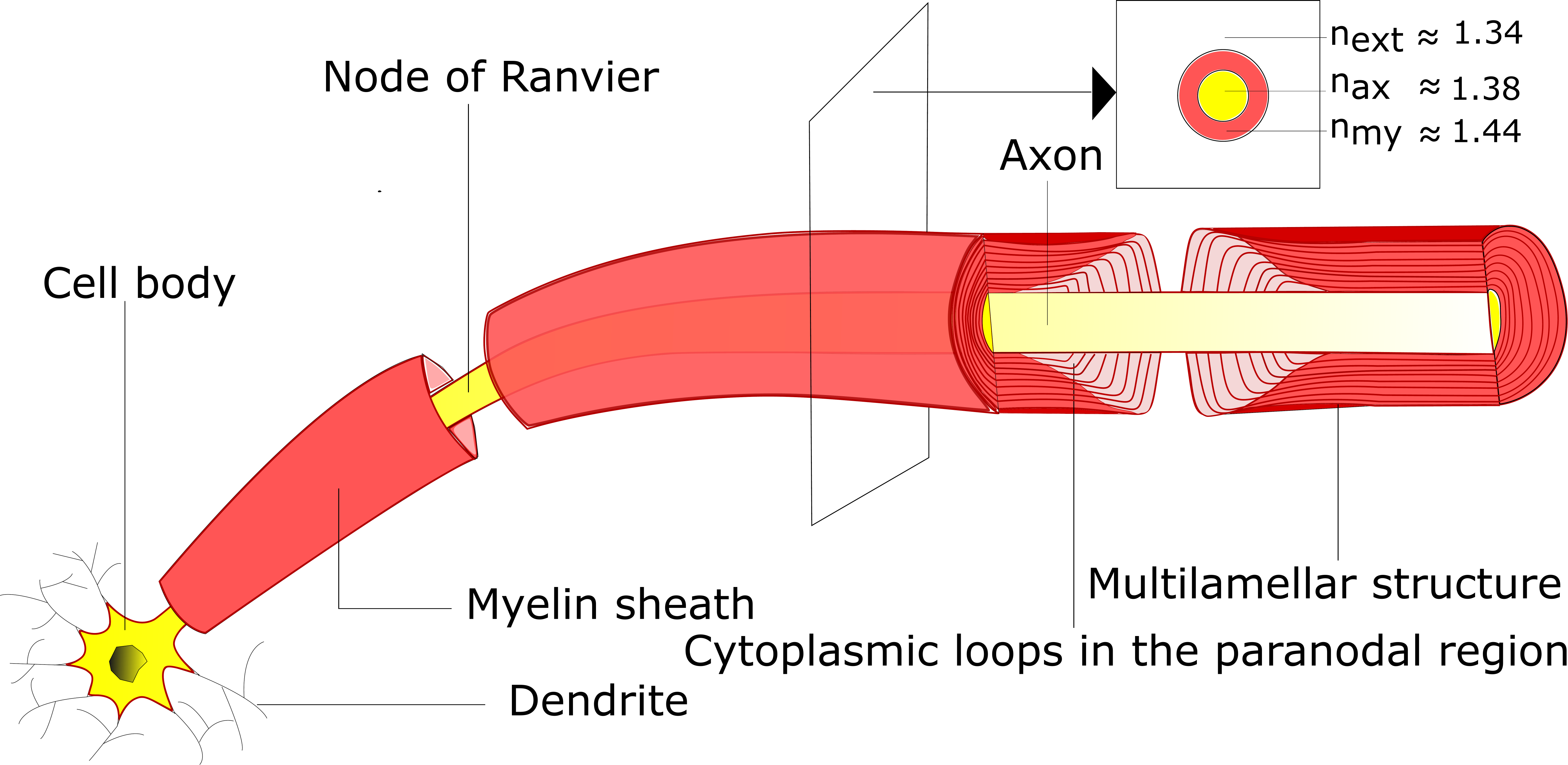}
\caption{\:\:\footnotesize{\textbf{3-D schematic representation of a segment of a neuron.} Different parts of a segment of a neuron whose myelinated axon is sliced longitudinally near the end of the segment to illustrate the structure better. The inset depicts the cross section in the transverse plane (perpendicular to the length) of the myelinated axon. Here $\mathrm{n_{my}}$, $\mathrm{n_{ax}}$, and $\mathrm{n_{ext}}$ are the refractive indices of the myelin sheath, the inside of the axon, and the interstitial fluid outside respectively. The compact myelin (shown in red) terminates near the Node of Ranvier, with each closely apposed layer of myelin ending in a cytoplasm filled loop (shown in light red) close to the axonal boundary.}}
\label{schematic}
\end{figure}

Here we propose a potential biophoton waveguide in the brain. Many axons are tightly wrapped by a lamellar structure called the myelin sheath, which has a higher refractive index \cite{antonov} than both the inside of the axon and the interstitial fluid outside (see Fig.\ \ref{schematic}). This compact structure could therefore also serve as a waveguide, in addition to increasing the propagation speed of an action potential (via saltatory conduction) based on its insulating property \cite{purves}. There is some indirect experimental evidence for light conduction by axons \cite{sun,tang1}, including the
observation of increased transmission along the axes of the white matter tracts, which consist of myelinated axons \cite{hebeda}. Myelin is formed in the central nervous system (CNS) by a kind of glia cell called oligodendrocyte. Interestingly, certain glia cells, known as Muller cells, have been shown to guide light in mammalian eyes \cite{franze, labin}. \par

An interesting feature of photonic communication channels is that they can transmit quantum information as well. The potential role of quantum effects in biological systems is currently being investigated in several areas, including olfaction \cite{turin,franco}, avian magnetoreception \cite{ritz,hiscock}, and photosynthesis \cite{engel,romero}. There is also growing speculation about the role of fundamental quantum features such as superposition and entanglement in certain higher level brain functions \cite{mashour,hameroff,jibu1,jibu2,fisher}. Of particular relevance is the ``binding problem'' of consciousness, which questions how a single integrated experience arises from the activities of individual molecules in billions of neurons. The answer to this question might be provided by quantum entanglement \cite{horodecki}, where the whole is more than the sum of its parts in a well-defined physical and mathematical sense.

The main challenge in envisioning a ``quantum brain'' is environmental decoherence, which destroys quantum effects very rapidly at room temperature for most physical degrees of freedom \cite{tegmark1}. However, nuclear spins can have coherence times of tens of milliseconds in the brain \cite{warren,lei}, and much longer times are imaginable \cite{fisher}. Long-lived nuclear spin entanglement has also been demonstrated in other condensed-matter systems at room temperature \cite{dolde}. A recent proposal on ``quantum cognition'' \cite{fisher} is based on nuclear spins, but relies on the physical transport of molecules to carry quantum information, which is very slow. In contrast, photons are well suited for transmitting quantum information over long distances, which is why currently envisioned man-made quantum networks rely on optical communication channels (typically optical fibers) between spins  \cite{kimble,sangouard}.

Efficient light guidance therefore seems necessary for both classical and quantum optical networks in the brain. Is this possible in myelinated axons with all their ``imperfections'' from a waveguide perspective? In an attempt to answer this question, we have developed a detailed theoretical model of light guidance in axons. We show in the next section that the answer seems to be in the affirmative.

\begin{figure*}
\includegraphics[scale=0.085]{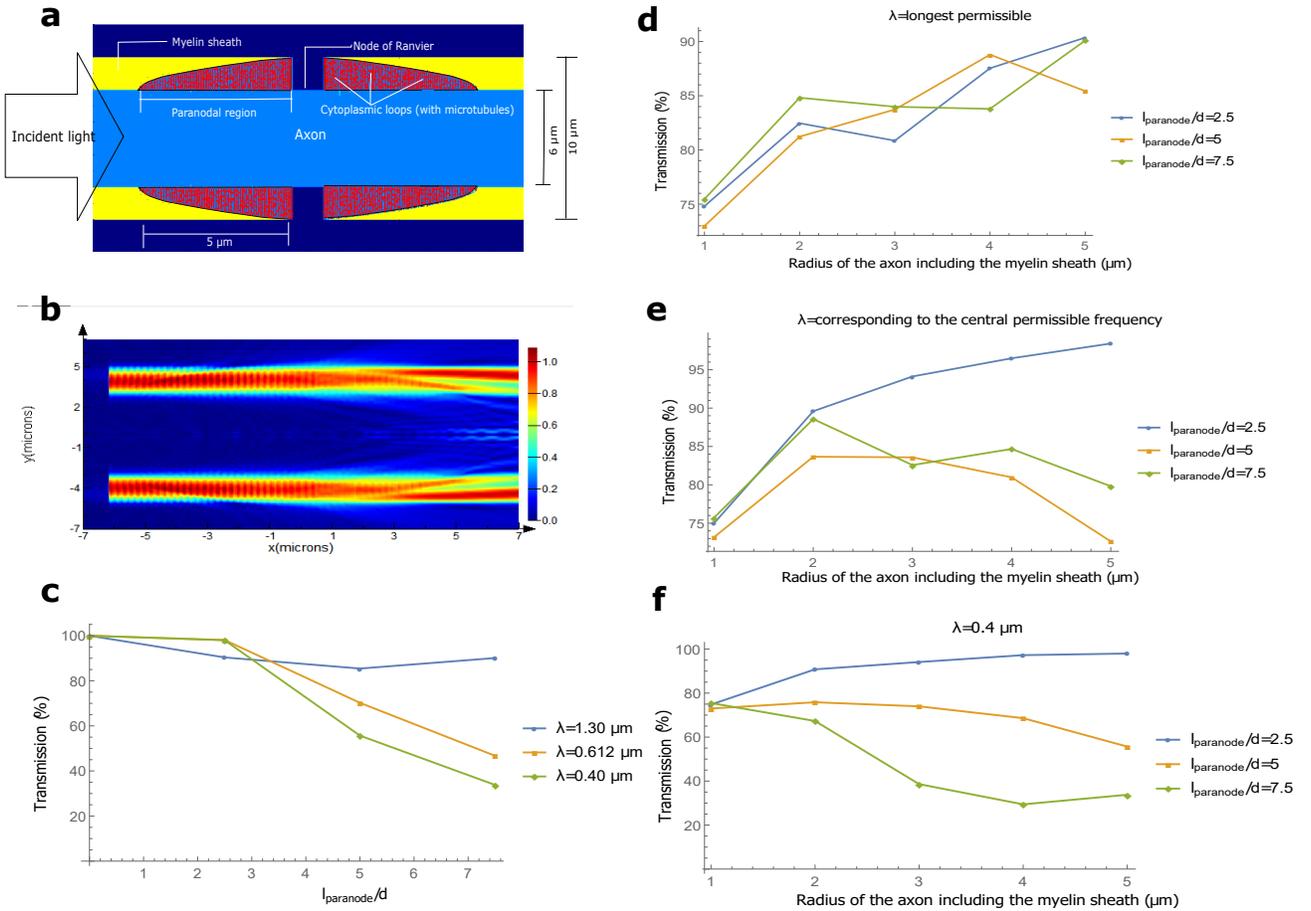}
\caption{\:\:\footnotesize{\textbf{Nodal and paranodal regions.} \textbf{(a)} Longitudinal cross-section of our 3-D model of the nodal and the paranodal regions. For this example, the radius of the axon including the myelin sheath, $r'$ = \SI{5}{\micro\meter}, and length of the paranode, $l_{paranode}$ = \SI{5}{\micro\meter}. \textbf{(b)} Magnitude of the electric field (in the frequency domain) in the longitudinal direction (EFPL) as a cylindrically symmetric input mode with wavelength \SI{0.612}{\micro\meter} crosses the region. \textbf{(c)} Transmission for an axon with $r'$ = \SI{5}{\micro\meter}, as a function of the \textit{p-ratio}, defined as $l_{paranode}/d$, where $l_{paranode}$ is the length of the paranodal region in one internode and $d$ is the thickness of the myelin sheath. \textbf{(d)}-\textbf{(f)} Transmission as a function of the axon caliber for different wavelengths and different paranodal lengths. The number of cytoplasmic loops, and the microtubules in the paranodal region are kept in proportion to the thickness of the myelin sheath, and the volume of the paranodal region respectively (see Methods).}}
\label{paranode}
\end{figure*}

\parskip = \baselineskip
\noindent
\textbf{\large{Results}}
\newline \textbf{Introduction to our approach}. We use Lumerical's software packages FDTD (Finite Difference Time Domain) Solutions and MODE Solutions for numerically solving the three dimensional electromagnetic field equations in various scenarios to elucidate the waveguiding characteristics of myelinated axons.  For the majority of our simulations, we take the refractive indices of the compact myelin sheath, the axon and the fluid outside to be 1.44, 1.38 and 1.34 respectively (see Fig.\ \ref{schematic}), consistent with their observed values \cite{antonov, wang, tuchin}. This index contrast allows guided modes of light inside the myelin sheath. Although there are many scatterers both inside the axon (cell organelles, e.g.\ mitochondria, and endoplasmic reticulum, lipid vesicles, and filamentous structures, e.g.\ microtubules, and neurofilaments) and outside (e.g.\ microglia, and astrocytes), modes confined primarily in the myelin sheath will effectively not see these structures. The modes are primarily confined in a waveguide if its dimension is close to or larger than the wavelength of the light. Myelinated axons in the brain greatly differ in length and caliber. The short axons of the interneurons are only $\sim$1 mm long, while the longest axons can run through almost the whole length of the brain with numerous branches. Their diameters range from 0.2 microns to close to 10 microns  \cite{liewald}. We shall assume the \textit{g-ratio} (the ratio of the radius of the axon, \textit{r} and the outer radius of the myelin sheath, \textit{r}$'$) equal to 0.6 for the majority of our simulations, close to the experimental average \cite{friede}.

Biophotons have been observed in the wavelength range \SI{0.2}{\micro\meter}--\SI{1.3}{\micro\meter}. Most proteins (including the proteins in the myelin sheath) strongly absorb at wavelengths close to \SI{0.3}{\micro\meter}. To avoid absorption, we shall consider the transmission of light with wavelengths above \SI{0.4}{\micro\meter}. For different axon calibers, we send in light at different wavelengths, ranging from \SI{0.4}{\micro\meter} to the thickness of the myelin sheath (denoted by $d$), or \SI{1.3}{\micro\meter} (the upper bound of the observed biophoton wavelength), whichever is smaller. This ensures good confinement in the myelin sheath to limit interactions with the inhomogeneous medium inside and outside the axon (see Supplementary Information). We call this upper wavelength bound the longest permissible wavelength ($\lambda_{max}$). The shortest permissible wavelength ($\lambda_{min}$) for all simulations is \SI{0.4}{\micro\meter}. In addition to $\lambda_{max}$, and $\lambda_{min}$, we choose an intermediate wavelength corresponding to the central permissible frequency (mid-frequency of the permissible frequency range), denoted by $\lambda_{int}$. In a single simulation, FDTD calculates the input mode at $\lambda_{int}$, and sends light at different wavelengths with the same spatial mode profile (see Methods). Note that for the thinnest axons considered, $\lambda_{max}$= $\lambda_{int}$= $\lambda_{min}$=\SI{0.4}{\micro\meter} (\textit{d}=\SI{0.4}{\micro\meter}, too, for good confinement).

Next, we  discuss the transmission of the guided modes of the structure (see Supplementary Information for different mode profiles) in the presence of the optical imperfections (e.g. discontinuities, bends and varying cross-sections). The theory of various imperfections in optical fibers is developed with long distance communication in mind, which requires very small imperfections, and focuses on the conventional fiber geometry, where the refractive index of the core is higher than that of the cladding. Since the myelin sheath based waveguide does not pertain to either of these conditions, we will be mostly dealing with explicit examples. We simulate short axonal segments as the computational resource requirements for FDTD are very high, and extrapolate the results for the full length of an axon.

\noindent
\textbf{Nodal, and paranodal regions}. The myelin sheath is interrupted at almost regular intervals by the `Nodes of Ranvier', leaving the axon bare for  approximately \SI{1}{\micro\meter} \cite{kandel}. The lamellae, whose fusion and apposition leads to the formation of the compact myelin, terminate near the nodes in the paranodal region such that each lamella ends in a loop filled with dense cytoplasm (see Fig.\ \ref{schematic}). Many of these cytoplasmic loops are attached to the axonal membrane. For a thin myelin sheath, the paradonal region is almost ordered, with the innermost lamella terminating first, and so on, but for thicker sheaths, many cytoplasmic loops do not reach the axonal surface, but terminate on other loops. The length of the  paranodal region should, however, depend roughly on the the number of myelin lamellae, and hence on the thickness of the myelin sheath. We shall call the ratio $l_{paranode}/d$, the \textit{p-ratio}, where $l_{paranode}$ is the length of the paranode in one internode (the axon between two consecutive nodes) and $d$ is the thickness of the myelin sheath as defined earlier; \textit{p-ratios} around 5 are realistic \cite{zagoren}.

Fig.\ \ref{paranode}a shows our model for the region of the axon close to the node (see Methods section for detailed discussion on the model), and Fig.\ \ref{paranode}b depicts the magnitude of the electric field in the longitudinal direction (along the length of the axon) in the frequency domain, as a cylindrically symmetric input mode crosses this region. We call this EFPL (Electric Field Profile in the Longitudinal direction). We vary the length of the paranodal region for an axon with $r'$=\SI{5}{\micro\meter} in Fig.\ \ref{paranode}c, and observe the corresponding modal transmission (power transmission in the guided modes) up to a wavelength away from myelin sheath boundaries (see Methods and Supplementary Information) for 3 different wavelengths, \SI{0.40}{\micro\meter}, \SI{0.61}{\micro\meter}, and \SI{1.30}{\micro\meter}. We interpret the results in terms of beam divergence and scattering, which are the two primary sources of loss here. Shorter wavelengths diverge less, but scatter more. For a short paranodal region (\textit{p-ratio}=2.5), shorter wavelengths have a higher transmission, but for longer paranodal regions, longer wavelengths fare better because scattering becomes a more powerful agent of loss than divergence as the length is increased. In general, the transmission drops for all the wavelengths as the \textit{p-ratio} increases, although the trend is less clear for the longest permissible wavelength.

In Fig.\ \ref{paranode}d--f, we simulate the nodal region for 5 different axon calibers, several different wavelengths, and different \textit{p--ratios}. In general, the greater the mode volume, the less is the divergence. So, a mode with a larger mode volume (corresponds to thicker myelin sheath) should diverge less for the same paranodal length. Here, however, we are dealing with ratios ($l_{paranode}/d$), rather than absolute values of the lengths, making intuition slightly difficult. Still, in Fig.\ \ref{paranode}d, we see that the most loosely confined modes ($\lambda_{max}$) crudely follow this intuition, and transmission increases for thicker axons. For a fixed axon caliber, the transmission does not depend on the paranodal length in a well-defined way. One possible explanation for this feature is the unconventional nature of the waveguide itself. The long wavelengths mostly suffer loss because of divergence. However, in these waveguides, not all the light that diverges is lost. There is a possibility of a fraction of the light diverging into the axon to come back into the myelin sheath at the end of the paranodal region. This sometimes even increases the transmission in the myelin sheath for longer paranodal regions.

In, Fig.\ \ref{paranode}e, and Fig.\ \ref{paranode}f, for \textit{p-ratio} = 2.5, the trend follows the intuition based on divergence. Increase in myelin thickness leads to better confinement, and less divergence. However, for larger \textit{p--ratios}, the trend almost reverses, and thicker axons perform worse than thin ones. In these cases, scattering becomes more relevant than divergence, and longer paranodal regions lead to greater scattering. \par

Even with such a sudden discontinuity in the sheath, we find that transmission can still be fairly high. To summarize, if the \textit{p-ratio} is small ($\sim$2.5), well confined modes (shorter wavelengths) yield higher transmission, whereas loosely confined modes fare better for larger \textit{p-ratios}. Thicker axons are usually better than the thinner ones for smaller \textit{p--ratios} ($\sim$2.5) at all wavelengths. However, for shorter wavelengths and larger \textit{p--ratios} ($\sim$5 or greater), thinner axons have higher transmission. We verified that the transmission after multiple paranodal regions can be approximately predicted by exponentiating the transmission through one (see Supplementary Information).

\begin{figure*}
\includegraphics[scale=0.23]{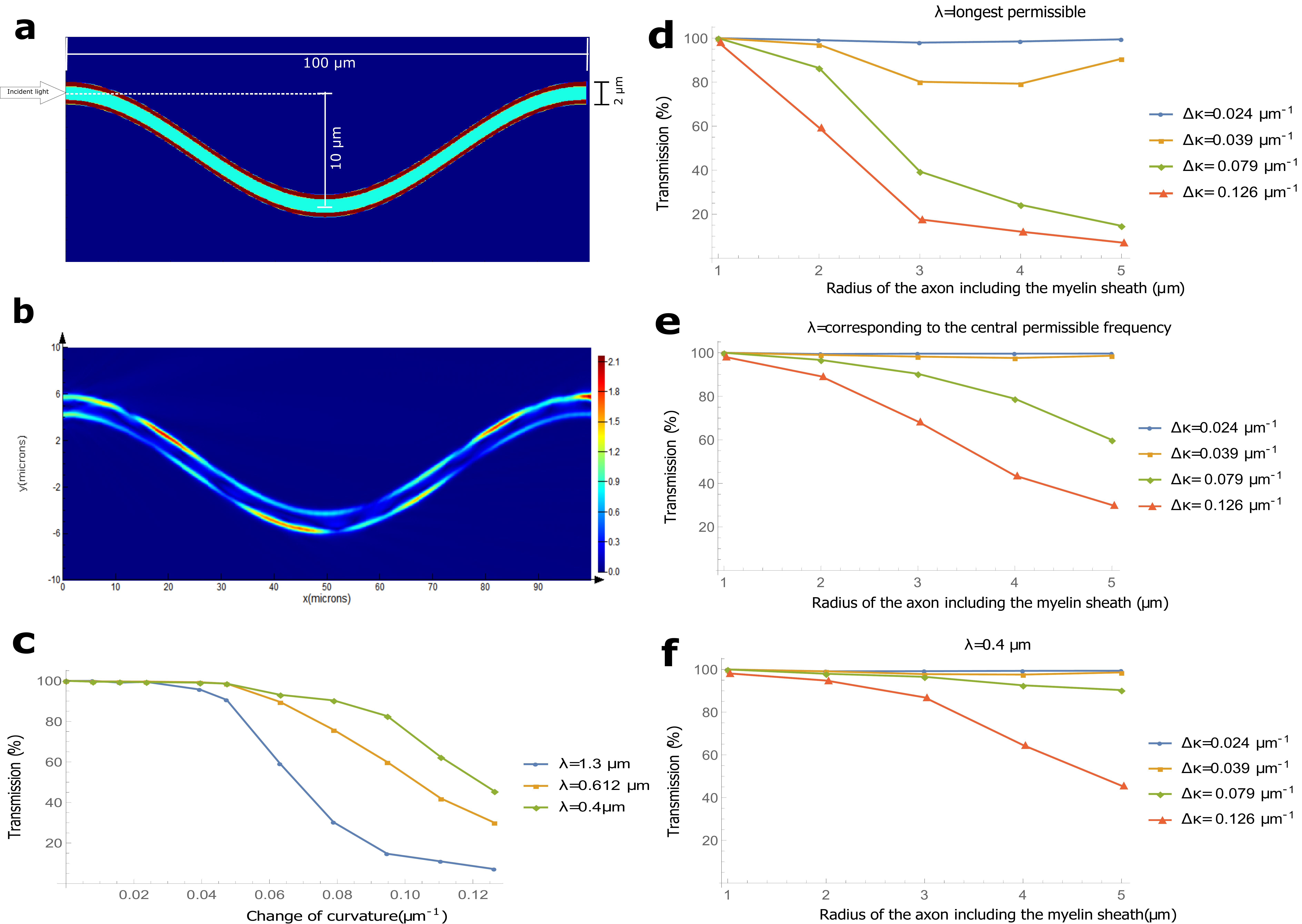}
\caption{\:\:\footnotesize{\textbf{Bends.} \textbf{(a)} The geometry of a sinusoidally bent waveguide. For this example, $r'$ = \SI{1}{\micro\meter}, and the amplitude (\textit{A}) and wavelength of the cosine function (\textit{l}) are \SI{5}{\micro\meter} and \SI{100}{\micro\meter} repectively. \textbf{(b)} EFPL as the input mode with wavelength \SI{0.4}{\micro\meter} crosses the region. \textbf{(c)} Transmission as a function of the change in curvature, $\Delta \kappa$, for different wavelengths in an axon with $r'$ =  \SI{5}{\micro\meter} ($\Delta \kappa$ is varied by varying $A$). \textbf{(d)}-\textbf{(f)} Transmission as a function of the axon caliber for different wavelengths and different $\Delta \kappa$.}}
\label{bend}
\end{figure*}

\begin{figure*}
\includegraphics[scale=0.23]{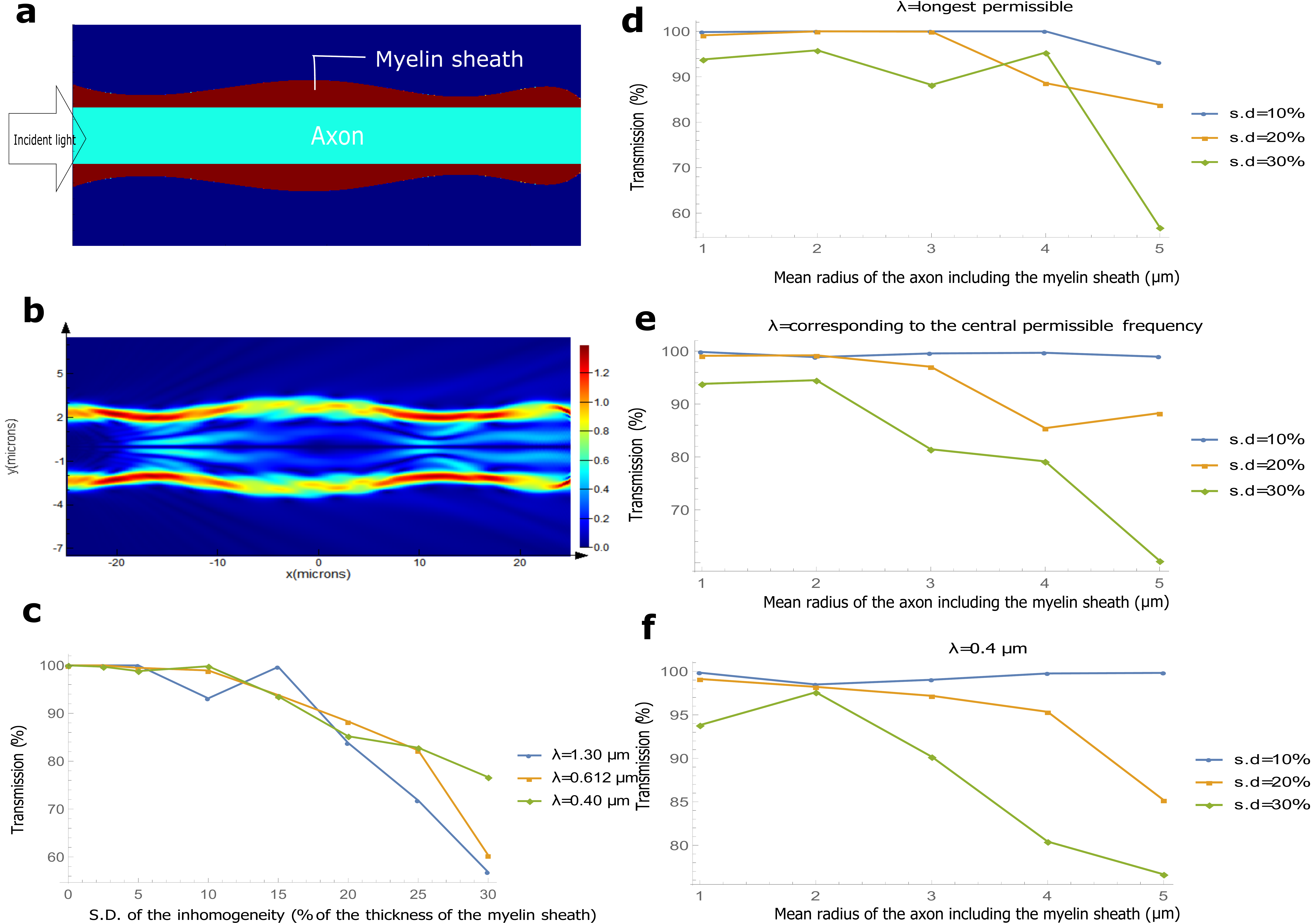}
\caption{\:\:\footnotesize{\textbf{Varying cross-sectional area.} \textbf{(a)} The geometry of a myelinated axon where the cross sectional area of the myelin sheath varies smoothly along the longitudinal direction. For this example, the mean radius of the axon with the myelin sheath is \SI{4}{\micro\meter} and the standard deviation (s.d.) of the variation of the myelin sheath's radius is \SI{0.48}{\micro\meter}. \textbf{(b)} EFPL as the input mode with wavelength \SI{0.612}{\micro\meter} crosses the region. \textbf{(c)} Transmission as a function of the s.d.\ of the variation in the myelin sheath's radius for different wavelengths (the mean radius of the axon with the myelin sheath is \SI{5}{\micro\meter}). \textbf{(d)}-\textbf{(f)} Transmission as a function of the mean radius of the axon including the myelin sheath for different wavelengths and different s.d.\ of the variation of myelin sheath's radius.}}
\label{ca}
\end{figure*}

\noindent
\textbf{Bends}. Optical power from the eigenmodes of a straight waveguide leaks out on encountering bends. Bends of constant curvature have eigenmodes which can propagate with minimal loss, but axons typically change their curvature along their length. These bent--modes (eigenmodes for circular bends) are more lossy than the straight--modes (eigenmodes for straight structures) for changing curvature. Therefore, an appropriate way to quantify the bend losses for an arbitrary axon path will be to incident the straight--mode in a waveguide with continuously varying curvature, and observe the transmission in the myelin sheath at the other end. We choose a sinusoidal waveguide since it has alternating regions of positive and negative curvatures, and can thus serve as a prototype for any arbitrary contour. Fig.\ \ref{bend}a is an example for an axon with radius \SI{0.6}{\micro\meter}, and Fig.\ \ref{bend}b shows the EFPL as a straight--mode passes through.
Bending losses for conventional S bend waveguides (half a cosine function) depend most strongly on the change of curvature \cite{syahriar}. We therefore plot total power transmission (calculated by integrating the real part of the Poynting vector of the output light directly across the required area, and dividing it by the source power) up to a wavelength away from the myelin sheath boundaries (see Methods and Supplementary Information) as a function of the change of curvature, $\Delta \kappa = 4 A k^2$  (\textit{k} is the wavenumber of the sinusoidal function) for 3 different wavelengths in Fig.\ \ref{bend}c ($r'$=\SI{5}{\micro\meter}). A shorter wavelength is better confined and therefore yields higher transmission. Fig.\ \ref{bend}d--f compare transmission for axons of different calibers. Note that we calculate $\Delta\kappa$ of the curve passing through the central axis of the axon. But the inner part of a bent axon has a larger curvature than the outer part at each point. Such a difference becomes particularly important for thicker axons, since they see a larger effective change of curvature than thinner axons, and suffer more loss for the same $\Delta \kappa$. For $\Delta \kappa$ $\sim$\SI{0.024}{\micro\meter^{-1}}, almost all the permissible wavelengths are guided with negligible loss for all axon calibers discussed. We assume that in a typical axon, regions of large curvature do not exist for considerable length (which seems justified \cite{schain}) and $\Delta \kappa$ is a good parameter to quantify the bend inhomogeneity. Some of the axonal segments (\SI{1}{\milli\meter}) in \cite{schain} appear relatively straight with $\Delta \kappa<$ \SI{0.05}{\micro\meter^{-1}}, which yields greater than 90 $\%$ transmission for thin axons.

\begin{figure*}
\includegraphics[scale=0.3]{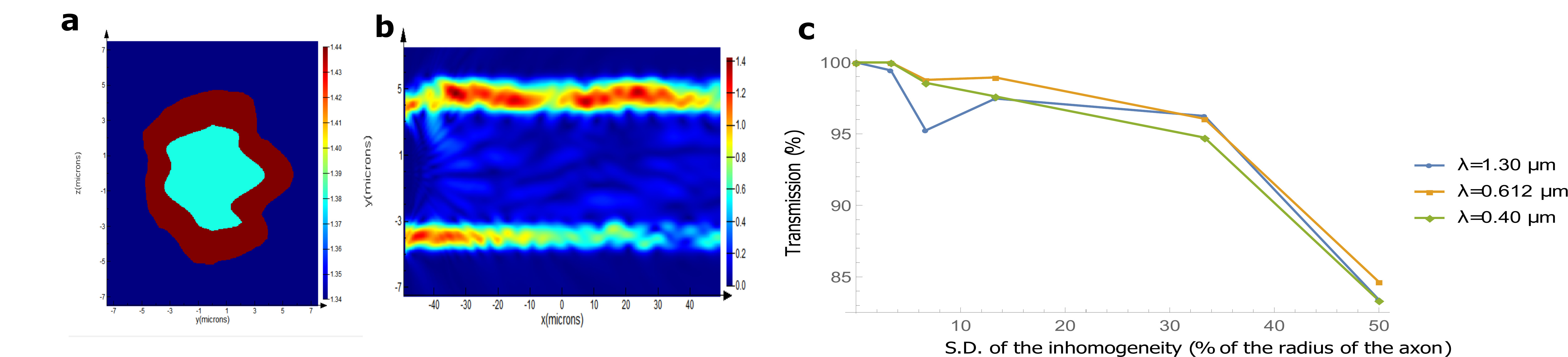}
\caption{\:\:\footnotesize{\textbf{Non-circular cross section of the axon and myelin sheath.} \textbf{(a)} An example of the cross-section of a myelinated axon. The mean distance of the points along the circumference of the axon from its center is \SI{3}{\micro\meter} and the s.d.\ is \SI{0.4}{\micro\meter}. The outer boundary of the myelin sheath is a parallel curve drawn at an approximate separation of \SI{2}{\micro\meter} from the axonal boundary. \textbf{(b)} EFPL as a cylindrically symmetric eigenmode for a circular cross-section ($\lambda$=\SI{0.612}{\micro\meter}, $r$=\SI{3}{\micro\meter}, and $r'$=\SI{5}{\micro\meter}) passes a straight waveguide with this non-circular cross-section. \textbf{(c)} Transmission as a function of the s.d.\ of the distance between the points on the circumference of the axon and a circle of radius \SI{3}{\micro\meter} for different wavelengths.}}
\label{nccs}
\end{figure*}

\noindent
\textbf{Varying cross-sectional area}.
The thickness of the myelin sheath is not uniform all along the length of the axon. We vary \textit{d} randomly according to a normal distribution. The mean of the distribution is in close agreement with that predicted by the \textit{g-ratio}, and the standard deviation (s.d.) of the distribution is varied. Fig.\ \ref{ca}a shows the longitudinal cross-section for one such simulation (\textit{r}=\SI{2.4}{\micro\meter}, length of the axonal segment is \SI{50}{\micro\meter}, and the s.d.\ is 30  $\%$ of the average thickness of the myelin sheath). Fig.\ \ref{ca}b shows the EFPL for input light with $\lambda = $ \SI{0.612}{\micro\meter}. In Fig.\ \ref{ca}c, we observe that, in general, a more random distribution of the radius suffers a greater loss (for all wavelengths), and shorter wavelengths transmit slightly better. Fig.\ \ref{ca}d--f compare the total power transmission (up to a wavelength from the myelin sheath boundaries) in axons of different calibers. Thinner axons can tolerate greater percentage-inhomogeneity, suggesting a closer dependence on the absolute value of the inhomogeneity. All the axons have close to unit efficiency in transmission for less than 10 $\%$ variation in radius. Extrapolation for transmission in a longer segment of the axon is straightforward. One can exponentiate the transmission fraction by the number of \SI{50}{\micro\meter} segments in the axon.  We have assumed a correlation length in the roughness of the myelin sheath boundary to be \SI{5}{\micro\meter}--\SI{10}{\micro\meter} (see Methods). Longer correlation lengths will yield better transmission for the same s.d. Some of the axonal segments (length $\sim$\SI{5}{\micro\meter}) of thin axons (\textit{r} $\sim$\SI{1}{\micro\meter}) are within this inhomogeneity, as seen in the images of \cite{peters}. We did not find suitable images of thicker myelinated axons, and longer segments from which a more realistic estimate of this particular inhomogeneity could be extracted.

\noindent
\textbf{Non-circular cross section}.
Axons can have quite arbitrary transverse cross-sectional shapes, and the ensheathing myelin partly imbibes that shape \cite{peters}. We give an example in Fig.\ \ref {nccs}a and the corresponding EFPL when an eigenmode for a circular cross-section ($\lambda$=\SI{0.612}{\micro\meter}, $r$=\SI{3}{\micro\meter}, and $r'$=\SI{5}{\micro\meter}) is incident on it in Fig.\ \ref{nccs}b. In this example, the points along the cross-sectional circumference of the axon are generated randomly according to a normal distribution  with a mean value \SI{3}{\micro\meter} and a standard deviation \SI{0.4}{\micro\meter} (13.33 $\%$ of the axon radius). The myelin sheath is an approximate parallel curve drawn at a perpendicular distance of \SI{2}{\micro\meter} (so that the average \textit{g-ratio} = 0.6) surrounding the axon. Fig.\ \ref{nccs}c shows the total power transmission (within a wavelength of the myelin sheath boundaries) in the myelin sheath for different shapes and different wavelengths in a \SI{100}{\micro\meter} long structure. As expected, transmission drops for all wavelengths as the cross-section becomes more random. Images in \cite{peters} show many axons with less than $10 \%$ inhomogeneity in the cross-sectional shape. If the axon and myelin sheath do not change the cross-sectional shape substantially along their length, there will be almost no more loss, as coupling loss is the primary source of loss here (rather than the propagation loss). Therefore, we do not attempt to calculate the effect of random cross-section for different axon calibers. However, if the cross-section changes significantly, there will be propagation loss as well (see Supplementary Information).

\noindent
\textbf{Other imperfections}.
In addition to the sources of loss discussed above, there can be several other imperfections, the most significant of which is the cross-talk between axons. Light in a myelinated axon would not leak out significantly, even if placed in direct contact with cells of lower refractive indices than the myelin sheath. However, if two or more myelinated axons are placed very close to each other (side by side), then light leaks out from one to the other (see Supplementary Information). Cross-talk can be interpreted both as a loss and a coupling mechanism between axons in a nerve fiber. In general, the axons should be a wavelength apart to prevent cross-talk, which seems realistic from some of the images in \cite{peters}.

The other imperfections that we considered do not affect the transmission significantly. The refractive indices of the axon, the myelin sheath, and the outside medium, were taken to be constants for the simulations so far. Next, we varied the refractive indices of the axon, and the myelin sheath, both transversely and longitudinally (with a correlation length for the random longitudinal variation of the refractive index $\sim$ \SI{5}{\micro\meter}--\SI{10}{\micro\meter}), keeping the mean the same as the one used so far, and a s.d.\ of 0.02 (typical variation as expected from \cite{antonov, wang}) for a few of the simulations . We observed no significant changes in the transmission (typically less than 1 $\%$). Moreover, there can be astrocytes and other glia cells in the nodal region close to the axon. As light crosses this region from one internode to the other, it will pass through these cells. We modeled them as spheres with radii varying from \SI{0.1}{\micro\meter} to \SI{0.3}{\micro\meter}, and refractive index 1.4, filling up one third of the volume of the nodal region outside the axon (expected from the images in \cite{peters}). The transmission increased slightly ($\sim$2 $\%$) for the thinnest axons, while it stayed almost unchanged for the thickest ones.

\noindent
\textbf{Absorption}.
In biological tissues, and more so in the brain, scattering of light, rather than absorption, is the main source of attenuation of optical signals \cite{cheong}. To our knowledge, the absorption coefficient of the myelin sheath has not been measured experimentally. We can only infer it indirectly with limited accuracy. The average absorption coefficient in the white matter decreases almost monotonically from $\sim$\SI{0.3}{\milli\meter^{-1}} to $\sim$\SI{0.07}{\milli\meter^{-1}} for wavelengths \SI{0.4}{\micro\meter} to \SI{1.1}{\micro\meter} \cite{yaroslavsky}. But myelin can not be responsible for the majority of the absorption since grey matter (almost devoid of myelin) has comparable absorption coefficients \cite{yaroslavsky}. It is likely that light sensitive structures (e.g. chromophores in the mitochondria) are the main contributors to the absorption. Another way to infer myelin's absorption coefficient is to look at the absorption of its constituents, i.e. lipids, proteins and water. Mammalian fat shows an absorption coefficient less than \SI{0.01}{\milli\meter^{-1}} for the biophotonic wavelength range \cite{veen}. Water has similar absorption coefficients. Most proteins have a strong resonance peak close to \SI{0.28}{\micro\meter} with almost negligible absorption above \SI{0.34}{\micro\meter}, and the proteins in the myelin (e.g. myelin protolipid protein, and myelin basic protein) behave similarly \cite{facci}. Thus, absorption in myelin for the biophotonic wavelengths seems negligible (over a length scale of $\sim$\SI{1}{\centi\meter}), based on the data of its constituents. Only a direct measurement could tell us more.

\noindent
\textbf{Attainable transmission}. We discuss a few examples to estimate the attainable overall transmission. The internodal length is typically equal to 100--150 times the axonal diameter \cite{friede,ibrahim}. For an axon with $r$ = \SI{3}{\micro\meter} ($r'$ = \SI{5}{\micro\meter}), internodal length = \SI{1}{\milli\meter}, wavelength of input light = \SI{1.3}{\micro\meter}, s.d.\ for varying area = 2.5 \%, $\Delta \kappa $ = \SI{0.039}{\micro\meter^{-1}}, s.d.\ for non-circularity in cross-section shape = 13.33 \%, separation from the nearby axons = \SI{1}{\micro\meter}, and \textit{p-ratio} = 7.5, the transmission after 1 cm would be $\sim$31 \% (see the Methods for the procedure). However, if the wavelength of input light = \SI{0.61}{\micro\meter}, $\textit{p-ratio}$ = 2.5, and all the other parameters are kept the same, the transmission could be $\sim$82 \%. A thinner axon with $r$ = \SI{1.8}{\micro\meter} ($r'$ = \SI{3}{\micro\meter}), internodal length = \SI{500}{\micro\meter}, wavelength of light = \SI{1.2}{\micro\meter}, $\textit{p-ratio}$ = 7.5, s.d.\ for varying area = 20 \%,  separation from other axons = \SI{1.2}{\micro\meter}, and $\Delta \kappa $ = \SI{0.039}{\micro\meter^{-1}} would yield $\sim$3 \% transmission after 1 cm. However, there are neurons in the brain whose axons are $\sim$\SI{1}{\milli\meter} long \cite{nolte} (e.g. the local interneurons). If we take a \SI{2}{\milli\meter} long axon, then the transmission for the 3 examples discussed above would be $\sim$78 \%, $\sim$96 \%, and $\sim$46 \% respectively.  The predominant loss for these examples is in the paranodal regions. Sources and receivers would need to be located close to the ends of the myelinated sections of the axon to reduce coupling losses. Let us note that photons could travel either way (from the axon terminal up to the axon hillock or the other way round) in an axon.

\noindent
\textbf{Attainable communication rates}. One potential challenge for the use of biophotons for inter-neuron communication is the fact that biophoton emission rates per neuron seem to be quite low. In \cite{tang1}, the authors count the number of biophotons emitted per minute by a slice of mouse brain, using a photodetector placed at a certain distance away, after exciting the neurons with glutamate, the most common excitatory neurotransmitter. Substituting the relevant experimental parameters, the estimated rate of biophoton emission is about 1 photon per neuron per minute. This estimate has significant uncertainty. On the one hand, the brain slice is strongly stimulated by glutamate, so the estimate might be high. On the other hand, only the scattered photons are counted. If there are photonic waveguides in the brain, most of the photons propagating in these waveguides would likely be absorbed in the brain itself rather than being scattered outside, so the estimate could also be much too low. It should also be noted that the emission rates could be very different depending on the specific neuron or neuron type.  Taking the above estimate at face value for the sake of the argument, such low photon rates could still be relevant. Given that there are about $10^{11}$ neurons in a human brain, there would still be over a billion photon emission events per second. This could be sufficient to transmit a large number of bits, or to distribute a large amount of quantum entanglement. In this context, it is worth keeping in mind that psychophysical experiments suggest that the bandwidth of conscious experience is less than 100 bits per second \cite{zimmermann,norretranders}. From a quantum perspective, it is known that the behavior of even moderate numbers of qubits (e.g. one hundred) is impossible to simulate efficiently with classical computers \cite{lloyd}.

\noindent
\textbf{Proposals to test the hypothesis}. There is some indirect evidence of light guidance in axons \cite{sun,tang1,hebeda}. As a way to test this \textit{in vitro}, we need to isolate a neuron with the necessary thickness of the myelin sheath, and small enough inhomogeneity, suspend it in a suitable solution to keep the cell alive for some time, and try to couple one of the guided modes into the axon. We could couple the light in close to the axon terminals, as real sources are suspected to be present there \cite{tang1}. To inject a guided mode in the myelin sheath and verify its guidance, one might need to decapitate the axon near the terminal and hillock regions, couple the mode directly in the myelin sheath, and observe the intensity (and if possible, the modal structure) of light emanating from the other end quickly, since the cellular properties start to change soon after death. Evanescent coupling and readout of light is another option.

For an \textit{in-vivo} test of light guidance, one might first try to prove the presence of photons in the myelin sheath. One could inject a light sensitive chemical (e.g.\ $\mathrm{AgNO_3}$) either in the cytoplasmic loops  in the paranodal region directly or in the oligodendrocytes, which would then circulate the chemical in the cytoplasmic loops, and possibly some to the myelin too. Light will activate the oxidation of $\mathrm{Ag^+}$ to Ag, which should be visible as dark insoluble granules. This is similar in spirit to the development of photographic films, and the in-situ biophoton autography (IBA) technique \cite{sun}.

Another interesting type of \textit{in-vivo} tests would involve the artificial introduction of sources and detectors into living neurons. Fluorescent molecules or nano-particles could serve as sources, and also as detectors, if their fluorescence can be triggered by the absorption of photons from the molecule or nano-particle that serves as the source \cite{yang}. An alternative possibility for the insertion of detectors may be provided by optogenetics \cite{deisseroth}, where specific kinds of neurons are genetically modified to produce proteins which can function as light sensitive ion-channels (e.g.\ channel rhodopsin). 
If we could embed these proteins specifically in the axonal membrane near a terminal end of the myelin sheath, or in the membranes of the cytoplasmic loops in the paranodal region at an end, and detect photons produced by an artificial source at the other end of the axon, we could verify the light guidance hypothesis. It is interesting to note that there is an increase in oligodendrogenesis and myelin sheath thickness near these genetically modified neurons when stimulated by light \cite{gibson}. Do the axons adapt themselves for better light guidance too (in addition to electrical guidance) by adding sufficient layers of myelin?

The final type of test would involve identifying naturally occurring sources and detectors in real neurons, and showing that photons are guided from the sources to the detectors. To our knowledge, photon emission has not yet been studied at the level of individual neurons. Photon measurements have been done macroscopically, counting only the scattered photons \cite{isojima, kobayashi, tang1, kataoka} (neglecting those which are guided or absorbed). It would be important to precisely pinpoint the sources of these photons and to characterize their wavelength and emission rates. This may be possible by enhancing the emission rates through nanoantennas \cite{kuhn}. It would also be very interesting to study the photon detection capabilities of potential natural detectors, such as centrosomes \cite{buehler} and chromophores in mitochondria \cite{kato}, ideally at the single-photon level. There may be other potential detectors that are yet to be discovered, e.g. light-sensitive proteins similar to channel rhodopsin used in optogenetics \cite{deisseroth}.

\noindent
\textbf{\large{Discussion}}
\newline We have shown that light conduction in a myelinated axon is possible even with realistic imperfections. We have proposed experiments to verify the key aspects of our hypothesis. We now briefly mention several related fundamental questions.

If photons are to serve as quantum communication links between nuclear spins, one also needs to explain how the photons and spins would interface with each other. Researchers in spin chemistry \cite{spin-chemistry} have discovered various ways in which electron and even nuclear spins can influence chemical reactions, which can also involve photons. A well-known biological example is provided by cryptochrome proteins, which can be activated by light to produce a pair of radicals with correlated electronic spins, which are suspected to be involved in bird magnetoreception (the ability to perceive magnetic fields) \cite{ritz}. Recent theoretical work suggests that interactions between electron and nuclear spins in cryptochromes are important for explaining the precision of the magnetoreception \cite{hiscock}. Cryptochromes are found in the eyes of mammals too (including humans), and they are also magnetosensitive at the molecular level \cite{foley}. Similar proteins, if  present in the inner brain regions, might act as an interface between biophotons and nuclear spins.

In order to connect individual quantum communication links to form a larger quantum network (allowing for the creation of entanglement between many distant spins), the nuclear spins interfacing with different axons would have to interact coherently, which might require close contact. The existence of synaptic junctions between individual axons is particularly interesting in this context.

Concerning the potential relevance of (classical or quantum) optical communication between neurons for consciousness and the binding problem, an interesting anatomical question would be whether brain regions that have been implicated in consciousness \cite{koch-review}, such as the claustrum \cite{koubeissi, crick}, the thalamus, hypothalamus and amygdala \cite{loewenstein}, or a recently identified ``hot zone'' in the posterior cerebral cortex \cite{koch-review} have myelinated axons with sufficient diameter to allow light guidance.


If optical communication along myelinated axons is indeed a reality, this would reveal a whole new aspect of the brain, with potential impacts on many fundamental questions in neuroscience.

\noindent
\textbf{\large{Methods}}
\newline{\footnotesize \textbf{Software packages}. We use Lumerical's FDTD Solutions, and Lumerical's MODE Solutions for all our simulations. Both these software packages use the Finite Difference Eigenmode (FDE) solver to generate the propagation modes for different waveguide geometries. FDE solves Maxwell's equations for the eigenmodes on a cross-sectional mesh using the finite difference algorithm \cite{zhu}. Finite Difference Time Domain (FDTD) method solves Maxwell's equations in time-domain on a discrete spatio-temporal grid formed by Yee cells \cite{yee}. Since FDTD is a time domain technique, it can cover a wide-frequency range in a single simulation. We use this feature to study the dependence of light guidance on the wavelength of the input light. But one has to be careful in the interpretation of the results. The two main areas of concern are the meshing accuracy of the simulation, and the change of the beam profile with wavelength. Lumerical's meshing algorithm refines the mesh for smaller wavelengths while leaves it coarse for the larger ones. We manually increase the mesh accuracy for all our large wavelength simulations to keep the number of Yee nodes almost constant for different wavelengths.  We have some tiny structures in our simulations (e.g. the microtubules), which need to be included in the mesh. We ensure that they are included by increasing the mesh accuracy to a setting such that the results converge for finer mesh. The variation of the beam profile with wavelength requires careful analysis too.  When we select a wide wavelength range, e.g. 400 nm to 1300 nm (equivalently 750 THz to 231 THz), and calculate the eigenmodes of the structure, FDTD calculates the eigenmode at the wavelength corresponding to the central  frequency, which is 612 nm (equivalently 490 THz) for this example. It injects light at different wavelengths but with the same spatial field profile. However, the mode-profiles for different wavelengths can differ substantially. Different kinds of waveguide imperfections need different analysis methods to account for this error, and we shall address this point individually for each one. We also ran multiple simulations (narrow sources at different wavelengths), where we send in the exact eigenmodes, and ensure that the results converge with that for a single simulation and a wide wavelength range.

\noindent
\textbf{Paranodal region}. The paranodal region is modelled by carving part of a paraboloid out of the cylinder comprising the axon and the myelin sheath, closely imitating their real geometry \cite{zagoren}. This part of the paraboloid is generated by revolving a segment of a parabola about the axis parallel to the length of the axon. This segment starts at the end of the paranodal region away from the node and terminates at the node. The general equation of the segment is $y=r+\sqrt{d^2/l_{paranode}\times x}$, where $r$, $d$, $l_{paranode}, x,$ and $y$ are the radius of the axon, the thickness of the myelin sheath, the paranodal length, the coordinate along the axis of the axon, and the coordinate perpendicular to the axis respectively. The paranodal region is divided into many cytoplasmic loops, modeled by the compartments between concentric rings of increasing radius as one approaches the node.  This is in accordance with the fact that the lamellae close to the axonal membrane terminates first and the most distant lamellae terminates last. The thickness of a ring is 10 nm, which is the typical thickness of the cell membrane. The number of these loops equals the number of the lamella in the compact myelin (average separation between adjacent lamellae is 20 nm \cite{lazzarini}). The microtubules in the cytoplasmic loops are generated randomly according to a uniform distribution, and placed transverse to the axon axis. They are concentric cylinders with inner and outer diameter equal to 13 nm and 6 nm respectively. The number of the microtubules is proportional to the volume of the paranodal region. The volume fraction of the microtubules (with respect to the paranodal region) is kept at 1.33 $\%$ which is a typical value of the volume fraction inside the axon \cite{yu}. The refractive indices of the cytoplasmic loops, the cell membrane, and the microtubules are taken to be 1.38, 1.50, and 1.50 respectively, close to their expected values \cite{sato,meyer}.

We obtain the electric field profile after a paranodal region, and expand it in an eigenbasis of the guided modes. Each time, we truncate this basis manually (for different axon calibers, and different wavelengths) in Lumerical's MODE Solutions, neglecting higher order modes (with effective refractive indices, $\mathrm{n_{eff}} <$ 1.34), almost all of which are lossy. Thus, our basis comprises of guided modes with $\mathrm{n_{eff}}$ between 1.44 and 1.34. The input mode is antisymmetric about two orthogonal axes in the cross-sectional plane (see Supp.\ Fig.\ 1). We label these axes Y and Z, with the origin at the center of the axon. Then $\mathrm{E_y}$ is antisymmetric about the Y axis, and $\mathrm{E_z}$ is antisymmetric about the Z axis, where $\mathrm{E_y}$, and $\mathrm{E_z}$ are the real parts of the Y and and Z components of the electric field E respectively.  Since the input mode is antisymmetric, and the structure is cylindrically symmetric, the guided modes will all be antisymmetric about the central axes. We therefore work in the subspace of antisymmetric modes, and expand the electric field profile in the basis of this subspace. Our waveguide permits a few guided modes primarily confined inside the axon. But the medium inside the axon (and outside the myelin) is expected to be scattering (see Supplementary Information), and we neglect the fraction of power residing in regions beyond a wavelength of the myelin sheath boundaries. We just integrate the real part of the Poynting vector (with the electromagnetic fields corresponding to the guided portion of the output light) across the myelin sheath  up to a wavelength from its boundaries (see Supplementary Information for the mathematical expressions). This is an approximate way of expanding the electric field profile in a basis of modes confined strongly in the myelin sheath. We choose to include the power within a wavelength of the myelin sheath boundaries to account for the evanescent fields and a few guided modes which are still very close to the myelin sheath, and are not strongly affected by the inhomogeneities inside and outside.

To account for the change in mode profile with wavelength, we expand the input mode (calculated at the central permissible frequency) in the basis of the guided modes at a particular wavelength. For shorter wavelengths, almost the entire power resides in a superposition of the guided modes (typically greater than 99.5 \%), but for longer wavelengths, the percentage of power in the guided modes can be significantly lower (e.g. for \textit{d} = \SI{2}{\micro\meter}, the expansion of an input mode in a basis of the modes at \SI{1.3}{\micro\meter} yield 97.11 \% coupling). So, we divide the output power (after the paranodal region) for longer wavelengths (obtained by integrating the real part of the Poynting vector with the electromagnetic fields corresponding to the guided portion of the output light across the myelin sheath  up to a wavelength from its boundaries) by the input power in the guided modes at those particular wavelengths (before the paranodal region) to obtain the normalized transmission.

\noindent
\textbf{Bends}.
Bends are generated by extruding a circular cross-section along a sinusoidal path. The cross-section is not exactly a circle, but a 26 sided polygon with the vertices lying on the corresponding radius (for the axon and the myelin sheath). All these vertices then follow the sinusoidal path to construct a bend. The path is discrete too, with a step size equal to \SI{0.5}{\micro\meter}. The number of vertices, and the step-size is optimized taking into account the accuracy and the speed of the simulation. With a straight path generated this way (discrete step size along the length, and a 26 sided polygon resembling a circle as the cross-section), and an eigenmode of the straight structure as the input, we ensure that we get close to 100 $\%$ transmission. The percentage transmission is calculated by integrating the real part of the Poynting vector (with the fields corresponding to the output light) across the myelin sheath up to a wavelength  away from the myelin sheath boundaries and dividing it by the source power. This is to include the evanescent fields and a few guided modes which are very close to the myelin sheath. Note that unlike the paranodal regions, we do not expand the output light in the basis of the guided modes at the end of the axon segment since the structure is continuously varying, and so is the basis of the guided modes. Some fraction of light in the non-guided modes at a particular cross-section might be included in the the basis of guided modes at an adjoining cross-section and vice-versa. Therefore it is more appropriate to observe the total power transmission instead of the modal transmission in such cases.  We continue to be cautious, and ignore all the power inside the axon (a wavelength away from the myelin sheath boundary).

To account for the difference in the mode-profiles at different wavelengths, we send in the eigenmode corresponding to the central permissible frequency in a uniform straight axon, and observe the transmission in the myelin sheath up to a wavelength. We observe that for the wavelength corresponding to the central permissible frequency and lower, the transmission is close to 100 $\%$.  But for longer wavelengths, the transmission can be substantially lower (e.g. for the thickest axon in our examples, the transmission is 96.81 $\%$ for the longest wavelength). If the right mode (corresponding to the longer wavelength) had been incident, we would have obtained 100 $\%$ transmission. To compensate for this insertion loss, we divide the transmission of the longer wavelengths by the transmission we obtain (for the same long wavelengths) when we send in a mode corresponding to the central frequency in a straight waveguide.

On rare occasions, for very small bends, this normalization procedure can yield slightly greater than 100 $\%$ transmission (the maximum observed overshoot was $\sim$0.18 $\%$) due to the finite resolution of the simulations, including the coarseness in the construction of the waveguide, and the import and export of field profiles across different programs (with different mesh sizes). In these cases the transmission is taken to be 100 $\%$.
We also adopt this approach for other inhomogeneities which face this overshoot problem.
We ran separate simulations with the exact input mode profiles for the particular wavelengths for a number of cases exhibiting the overshoot issue to verify that the transmission is indeed very close to unity in these cases.

We verify that the change of curvature seems to be the most important loss factor in the case of bends by running a few simulations for longer axonal segments (\SI{150}{\micro\meter}). For the same $\Delta \kappa$, the transmission for the longer segments was comparable to the transmission for the shorter ones.

\noindent
\textbf{Variable cross sectional area}.
The cross-sectional area of the myelin sheath is varied randomly according to an approximate normal distribution. We first generate 11 random points along a \SI{50}{\micro\meter} segment with the desired mean and the standard deviation (s.d.) in Mathematica (assuming a correlation length $\sim$ \SI{5}{\micro\meter}--\SI{10}{\micro\meter}). As an example, for an axon with \textit{r} = \SI{3}{\micro\meter}, \textit{d} = \SI{2}{\micro\meter}, and s.d.\  10 $\%$ of \textit{d}, the mean and s.d.\ of the points generated are \SI{5}{\micro\meter} and \SI{0.2}{\micro\meter} respectively. We fit these points with a polynomial of degree 7 (optimized over several trials). A polynomial of order 10 fits all the points exactly, but the local extrema of the function usually extend outside the span of the points it connects, and thus it has a greater randomness than that of the generated points. To ensure that the points in the fitted function indeed follow this distribution (with the expected mean and s.d.), we calculate the mean and the s.d.\ of this function by evaluating it at 200 points in the \SI{50}{\micro\meter} segment. This process is repeated many times to get an appropriate function  with the s.d.\ within 2.5 $\%$ of the desired value. The Gaussian nature of the randomness of the function is ensured manually (by observing that $\sim$95 $\%$ of the points lie inside 2 s.d.). The normalization to account for the change of mode profiles with wavelength is exactly the same as that for the bends.

\noindent
\textbf{Non-circular cross-section}.
The non-circular cross-section (in the X-Y plane) is generated, similar in spirit to the varying cross-sectional area. Here, the random points, corresponding to the vertices of the cross-section of the axon, are generated separately for the 2 halves (one in the positive Y plane and one in the negative Y plane). In the positive Y plane, 10 points are generated at equal intervals from the polar angle 0 to $\mathrm{\pi}$ such that the mean separation of these points from the center is kept constant (equal to the mean radius). Now, a polynomial of order 7 is fitted to these points. The 2 points  where this function crosses the X-axis (corresponding to the polar angles 0 and $\mathrm{\pi}$) are noted. In the negative  Y plane, 8 random points are generated at different polar angles. The other 2 points are those where the former function crossed the X axis. Now, a second fitting function (polynomial of degree 7) is generated with the weights of the couple of points lying on the X axis kept high to ensure that the function passes through these points. This is required because we want a continuity in the cross-sectional boundary for both the halves. This is the procedure for the construction of the cross-section of the axon. For the myelin sheath, we need to generate a parallel curve ensheathing the axon at a fixed perpendicular distance from the boundary of the axon. But a unique parallel curve for the  \textit{g-ratio} = 0.6 exists only when the s.d.\ of the boundary of the axon is small. For larger s.d.\, the segments in the generated parallel parametric curve start intersecting. Only an approximate parallel curve can be drawn in this case. We do that manually by selecting $\sim$50 points separated from the axon's boundary at the required perpendicular distance. Thus, the myelin sheath boundary is actually a $\sim$50 sided polygon.

We choose a relatively long axonal segment (\SI{100}{\micro\meter}) and verify that almost all the non-guided modes of the waveguide are lost during propagation (by noting the transmission across many different cross-sections along the length and seeing that they converge). We integrate the real part of the Poynting vector just across the myelin sheath (not up to a wavelength) for each wavelength. We divide this transmission by the transmission just in the myelin sheath for the corresponding wavelength in a straight cylindrical waveguide of the same length, when the cylindrically symmetric mode (eigen mode for a circular cross-section) corresponding to the central permissible frequency is incident. This gives us an approximate normalized transmission for each wavelength. Following the procedure adopted while dealing with the inhomogeneities discussed earlier (e.g. bends, and varying cross-sectional area), we could have constructed separate inner and outer parallel curves at a wavelength separation from the myelin sheath and integrated the real part of the Poynting vector across that region. However, such unique parallel curves do not exist for long wavelengths and large inhomogeneities, and drawing approximate curves manually would also yield only approximately correct transmission values. We have verified that the results obtained using both procedures almost match with each other for a number of trial cases (within $\sim$2 $\%$ of each other). Since the transmission under such an inhomogeneity (the cross-sectional shape remains the same) almost does not drop with further increase in axonal length, slight inaccuracies in the transmission values do not matter.

\noindent
\textbf{Procedure for estimating the attainable transmission}.
We considered several optical imperfections to estimate the attainable transmission over the total length of an axon. We exponentiate the transmission fraction (obtained in our simulations for short axonal segments) for the variable cross-sectional area, the nodal and the paranodal regions, and the cross-talk between axons the required number of times. We do not exponentiate the transmission fraction for bends and non-circular cross-sections. For bends, as discussed earlier, we believe that the transmission depends primarily on the change of curvature (irrespective of the total length). For non-circular cross-sections, all the loss can considered as coupling loss (propagation loss is negligible). We then multiply all these transmission fractions to obtain the net transmission over the total length of an axon.

\normalsize
\noindent
\textbf{Acknowledgements}
\newline
\footnotesize We thank J. Beggs, D. Bouwmeester, C. Brideau, M. Cifra, P. Colarusso, T. Craddock, J. Dai, J. Davidsen, A. Faraon, N. Forde, S. Hastings-Simon, P. King, J. Moncreiff, G. Popescu, N. Singh, P. Stys, and M. Tittel-Elmer for useful discussions and helpful comments, H. Jayakumar and the Lumerical technical support team for assistance with Lumerical's FDTD and MODE Solutions software packages, and H. W. Lau, H. Kaviani, C. Healey, the University of Calgary IT support team, and the Westgrid support team for assistance with running FDTD on a computer cluster.

\normalsize
\noindent
\textbf{Author contributions}
\newline
\footnotesize C.S. conceived the project with help from J.T.; S.K. and K.B. performed the calculations with guidance from P.B. and C.S.; S.K. and C.S. wrote the paper with feedback from all other co-authors.

\normalsize
\noindent
\textbf{Additional Information}
\newline
\footnotesize {\bf Competing financial interests}. The authors declare no competing financial interests.

\normalsize

\pagebreak
\clearpage
\widetext
\begin{center}
\textbf{\large Supplementary Information for ``Possible existence of optical communication channels in the brain''}
\end{center}
\setcounter{equation}{0}
\setcounter{figure}{0}
\setcounter{table}{0}
\setcounter{page}{1}
\makeatletter
\renewcommand{\theequation}{S\arabic{equation}}
\renewcommand{\thefigure}{S\arabic{figure}}
\renewcommand{\bibnumfmt}[1]{[S#1]}
\renewcommand{\citenumfont}[1]{S#1}

\renewcommand{\figurename}{Supplementary Figure}

\begin{figure}[H]
\begin{center}
\includegraphics[scale=0.08]{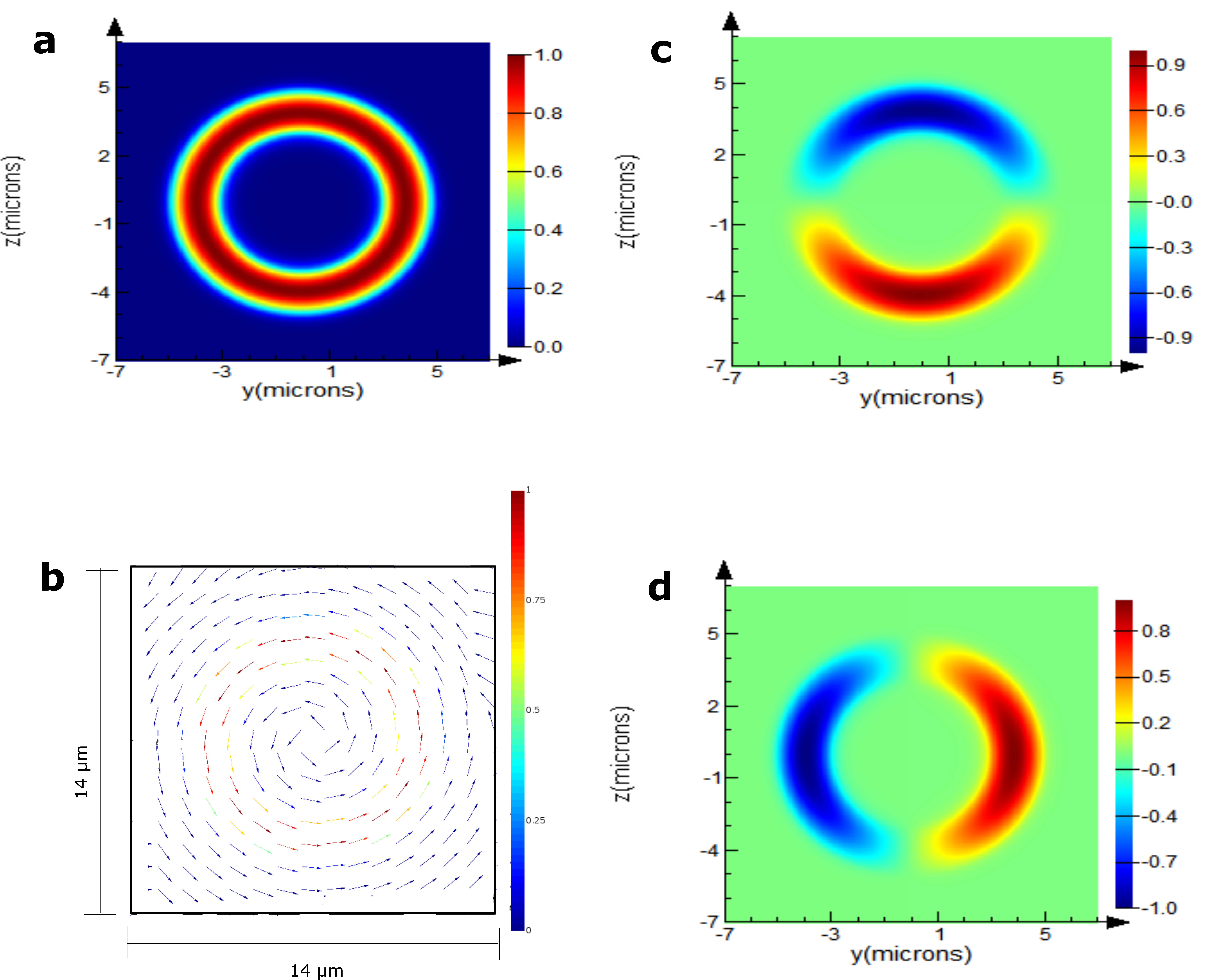}
\caption{\:\: \footnotesize{\textbf{Input mode} (a) Magnitude of the electric field of the input mode for a myelinated axon with the inner and outer radii of the myelin sheath, \textit{r} and \textit{r}$'$, as \SI{3}{\micro\meter} and \SI{5}{\micro\meter} respectively ($\mathrm{\lambda}$ = \SI{0.612}{\micro\meter}). (b) A vector plot of the electric field showing the azimuthal polarization of the input mode.  For clarity in the depiction of the direction of the field at different points, the arrow length is renormalised to the same value everywhere. The color bar on the side depicts the actual field magnitude.    (c)-(d) Electric field component along the Y direction ($E_y$). (d) Electric field component along the Z direction ($E_z$).}}
\label{inputmode}
\end{center}
\end{figure}

\noindent
\textbf{Guided modes}. An ideal waveguide allows electromagnetic waves with specific spatial field profiles to propagate without loss. These field profiles are the guided modes of the waveguide. Let us pick the thickest axon in our analysis (radius of the axon, \textit{r} and outer radius of the myelin sheath, \textit{r}$'$ are \SI{3}{\micro\meter} and \SI{5}{\micro\meter} respectively) to explain a few relevant details associated with these modes.  Supp.\ Fig.\ \ref{inputmode}a shows the power distribution of a cylindrically symmetric eigenmode of this structure for the wavelength \SI{0.612}{\micro\meter}, obtained using the Finite Difference Eigenmode (FDE) solver in the software Lumerical's FDTD Solutions. The electric field is azimuthally polarized (see  Supp.\ Fig.\ \ref{inputmode}b) to prevent modal dispersion in the birefringent myelin sheath, whose optic axes point in the radial direction \cite{chinn_S}. In the ray picture, this corresponds to `ordinary rays'. This is a Transverse Electric (TE) mode; the electric field oscillates in a plane transverse to the direction of propagation. It is similar to the $\mathrm{TE_{01}}$ mode of a conventional fiber \cite{jocher_S} in its spatial configuration of the field direction, i.e.\ both are azimuthally polarized. For a perfect waveguide, this mode will be guided without dispersion (because of birefringence) or other losses indefinitely. We have hundreds of other modes for this thickness of myelin sheath. Photons generated by a realistic source in the axons could couple to all these modes, with various coupling coefficients. Here, however, for the sake of simplicity (and lack of knowledge of realistic photon emission characteristics by potential sources), we start with a single mode and study the transmission of power in all the guided modes. The exact analytic form for the guided modes would involve linear combinations of  different Bessel functions, similar to those in \cite{kawakami_S}. However, we can come up with much simpler approximate expressions of the mode profiles observing those generated by the software. Supp.\ Fig.\ \ref{inputmode}a has a radial intensity dependence that is very close to a Gaussian, with peak intensity at the center of the myelin sheath, and with continuously decreasing intensity on both sides. The beam diameter corresponds to some fraction of the thickness of the myelin sheath (intensity of the form $Ae^{-(r-r_0)^2/(2  \sigma^2)}$, where $A$, $r$, $r_0$, and $4 \sigma$ are the amplitude, radial coordinate, the radial distance of the center of the myelin sheath, and the beam diameter respectively). The fraction can be estimated by knowing the fraction of the optical power inside the myelin sheath (e.g. 95.4 \% power in the myelin would imply that $4 \sigma = d $, where \textit{d} is the myelin sheath thickness). Note that this discussion about the approximate Gaussian shape of the field intensity is just to provide an intuition about the modes. In all our simulations, we use the modes directly generated by the software, and not the ones based on these simple approximate expressions.

In Supp.\ Fig.\ \ref{modeconfinetable}, we tabulate the modal fraction (fraction of the total power of a mode) inside the myelin sheath for different axon calibers and different wavelengths to illustrate their confinement. The power confined in the myelin sheath varies from 99.58 $\%$ for the best confined mode in the thickest axon in our simulations to 82.13 $\%$ for the least confined mode in the thinnest one, which is still higher than the typical confinement in the core of practical single mode fibers \cite{sairam_S} used for communication over tens of kilometres. Good confinement is necessary to limit interactions with the inhomogeneous medium inside and outside the axon. The scatterers inside the axon are the cell organelles, e.g.\ mitochondria, microtubules, and neurofilaments, whereas on the outside there are different types of cells, e.g.\ microglia, and astrocytes. There are guided modes with much weaker power confinement in the myelin sheath (less than 50 $\%$). However they might soon be lost to the inhomogeneities, and are therefore neglected. Supp.\ Fig.\ \ref{modeconfinetable} also explicitly lists the thickness of the myelin sheath (\textit{d}), the longest permissible wavelength ($\lambda_{max}$), the wavelength corresponding to the central permissible frequency ($\lambda_{int}$), and the shortest wavelength ($\lambda_{min}$) for each axon caliber. To remind the readers, for different axon calibers, we send in light at different wavelengths, ranging from \SI{0.4}{\micro\meter} (chosen to avoid absorption by the proteins) to the thickness of the myelin sheath, or \SI{1.3}{\micro\meter} (the upper bound of the observed biophoton wavelength), whichever is smaller for good confinement in the myelin sheath (at least 80 $\%$). We call this upper wavelength bound the longest permissible wavelength ($\lambda_{max}$). The shortest permissible wavelength ($\lambda_{min}$) for all simulations is \SI{0.4}{\micro\meter}. In addition to $\lambda_{max}$, and $\lambda_{min}$, we choose an intermediate wavelength corresponding to the central permissible frequency (mid-frequency of the permissible frequency range), denoted by $\lambda_{int}$. In a single simulation, FDTD calculates the input mode at $\lambda_{int}$ and sends light at different wavelengths with the same spatial mode profile. Note that for the thinnest axons considered, $\lambda_{max}$= $\lambda_{int}$= $\lambda_{min}$=\SI{0.4}{\micro\meter} (\textit{d}=\SI{0.4}{\micro\meter}, too, for good confinement).

\begin{figure*}
\begin{center}
\includegraphics[scale=0.70]{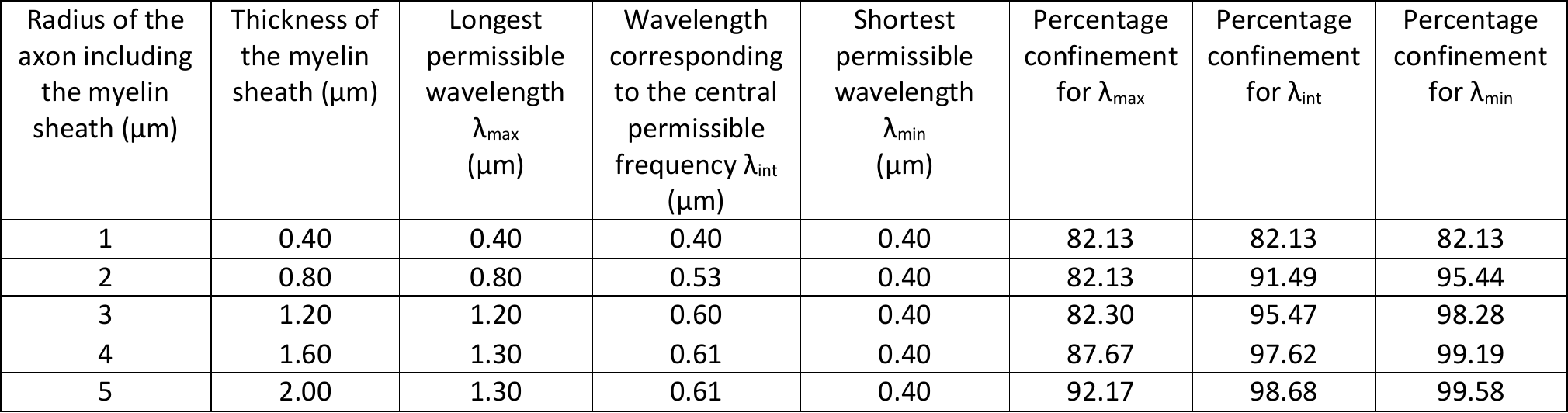}
\caption{\:\: \footnotesize{\textbf{Modal confinement in the myelin sheath.}  Range of permissible wavelengths for different myelin thicknesses and the percentage of power confined in the myelin sheath for those wavelengths.}}
\label{modeconfinetable}
\end{center}
\end{figure*}

\begin{figure*}
\includegraphics[scale=0.4]{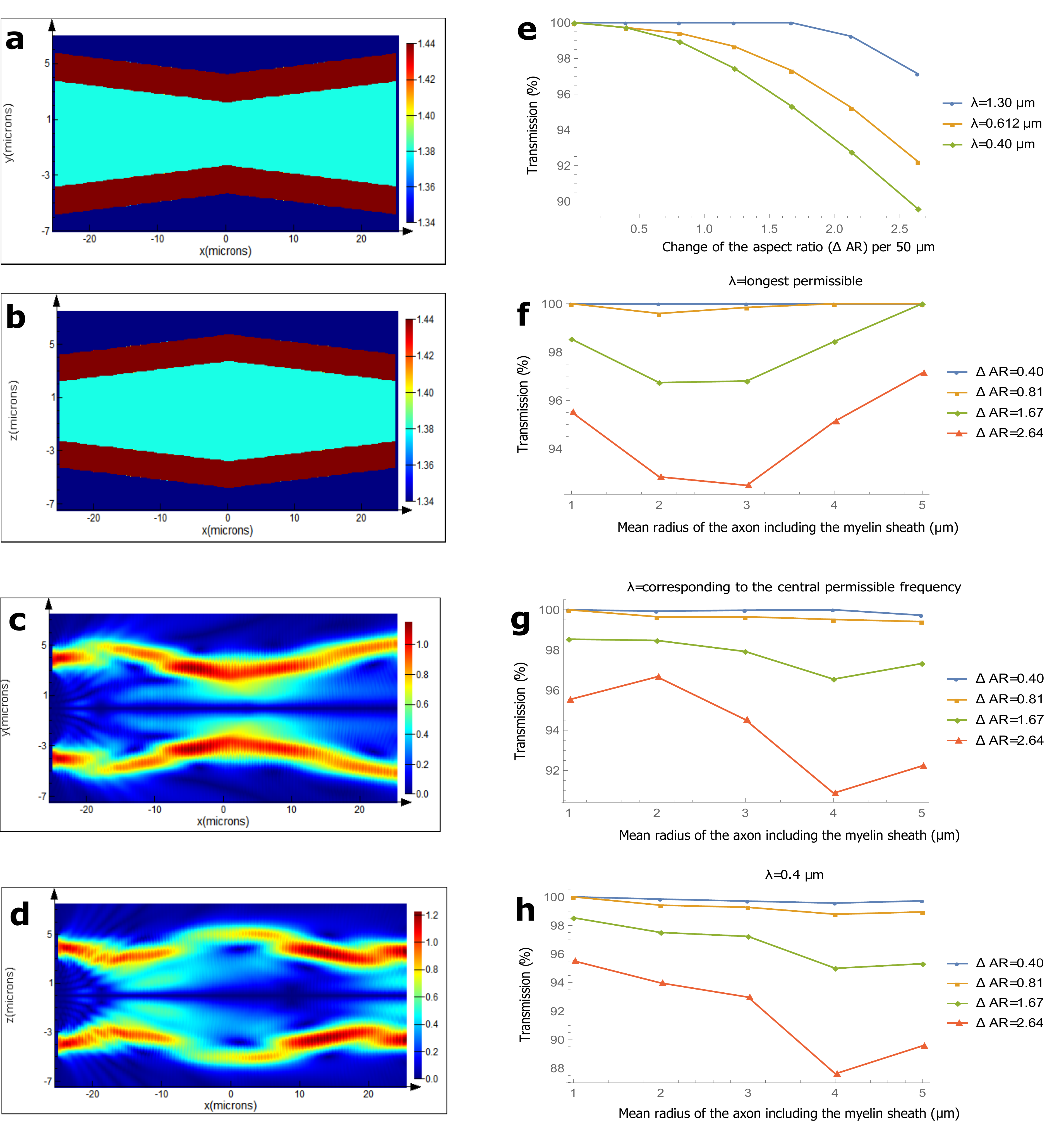}
\caption{\:\: \footnotesize{\textbf{Continuously varying non-circular cross-sectional shape.} (a)-(b) The refractive index profile of a myelinated axon in the X-Y plane and the X-Z plane respectively. The semi-major and semi-minor axes of the ellipses denoting the axonal boundaries at the start of the segment are \SI{3.75}{\micro\meter}, and \SI{2.25}{\micro\meter} respectively (the corresponding axes for the myelin sheath's outer boundaries are \SI{5.75}{\micro\meter},  and \SI{4.25}{\micro\meter} respectively).  (c)-(d)  Magnitude of the electric field (in the frequency domain) in the longitudinal direction (EFPL), as an eigenmode of a cylindrical waveguide ($r$ = \SI{3}{\micro\meter}, $r'$ = \SI{5}{\micro\meter}, and $\lambda$ = \SI{1.3}{\micro\meter}) crosses the axonal segment in the X-Y plane and the X-Z plane respectively. (e) Transmission as a function of the change in the aspect ratio ($\Delta AR$); $\Delta AR$ is defined as change in the ratio of the axes of the ellipse along two fixed orthogonal directions (here the Y and Z axes). The mean of the semi-axes of the axonal ellipse is \SI{3}{\micro\meter} (corresponding mean for the myelin sheath's outer boundary is \SI{5}{\micro\meter}). (f)-(h) Transmission as a function of the axon caliber for different wavelengths and different $\Delta AR$.}}
\label{nccsc}
\end{figure*}

Next, we shall discuss effects of a few imperfections in detail, expanding on the points mentioned in the main text.

\parskip = \baselineskip
\noindent
\textbf{Continuously varying non-circular cross-sectional shape}.
The cross-sectional shape of an axon changes in the longitudinal direction. In our model, we twist an axon, such that it starts out with an elliptical cross-section with semi-major and semi-minor axes \textit{a} and \textit{b} respectively, interchange  the axes midway (\SI{25}{\micro\meter}) and revert to their original shape at the end of the segment (\SI{50}{\micro\meter}). Since the cross-section is continuously changing, the guided modes at each section change too. An appropriate way to quantify the loss in such a structure would be to incident a cylindrically symmetric mode (identical to Fig.\ \ref{inputmode}) and observe its transmission at the other end.  Supp.\ Fig.\ \ref{nccsc}a, and Supp.\ Fig.\ \ref{nccsc}b show the longitudinal cross-section of the structure in 2 different planes (here, the X-Y and the X-Z planes). Supp.\ Fig.\ \ref{nccsc}c--d depict the magnitude of the electric field (in the frequency domain) along the length of an axon in those planes, as an eigenmode of a cylindrical waveguide ($r$ = \SI{3}{\micro\meter}, $r'$ = \SI{5}{\micro\meter}, and $\lambda$ = \SI{1.3}{\micro\meter}) passes by. We call this EFPL (Electric Field Profile in the Longitudinal direction). Supp.\ Fig.\ \ref{nccsc}e shows the total power transmission (calculated by integrating the real part of the Poynting vector of the output light directly across the required area, and dividing it by the source power) upto a wavelength away from the myelin sheath boundaries, as a function of the change in the aspect ratio (defined as the change in the ratio of the axes of the ellipse along two fixed orthogonal directions, here the Y and Z axes) of the ellipse per \SI{50}{\micro\meter}. We notice that longer wavelengths transmit better. We see transmission as a function of axon caliber in Supp.\ Fig.\ \ref{nccsc}f--h. Supp.\ Fig.\ \ref{nccsc}f, dealing with transmission for the longest permissible wavelengths, shows an interesting dip in transmission for \textit{r}$'$=\SI{2}{\micro\meter}, and \textit{r}$'$=\SI{3}{\micro\meter}. Comparing the transmissions for certain axon caliber (e.g. \textit{r}$'$=\SI{2}{\micro\meter}, and \textit{r}$'$=\SI{3}{\micro\meter}), and different wavelengths in Supp.\ Fig.\ \ref{nccsc}f--h, we observe that the intermediate wavelength has a larger transmission. We  note from Supp.\ Fig. \ref{modeconfinetable} that $\lambda_{max}=d$ for them, while for thicker myelin sheaths, i.e.  \textit{r}$'$=\SI{4}{\micro\meter}, and \textit{r}$'$=\SI{5}{\micro\meter}, $\lambda_{max}<d$. These observations suggest that there is an intermediate wavelength somewhere between \textit{d} and $\lambda_{min}$ (not necessarily  $\lambda_{int}$) where transmission is maximized. The propagation loss can be understood as a coupling loss between subsequent cross-sections (infinitesimally apart from each other). Shorter wavelengths have a higher number of guided modes at each cross-section than longer wavelengths, but the input mode at a shorter wavelength can get distorted more too (by exciting  higher-order modes). If it is distorted beyond a certain extent, light in those higher-order modes would be lost in subsequent cross-sections that do not have similar modes. Or if these higher order modes are at a wavelength away from the myelin sheath boundaries, they are not included in the transmission. So there is a competition between the number of available modes to couple to, and the extent of distortion. An intermediate wavelength turns out to be optimum.  Also, for larger $\Delta AR$, and short wavelengths, thinner axons are better, suggesting the relevance of the absolute value of the change in the ellipse's axes. The transmission for close to $\Delta AR$ (per \SI{50}{\micro\meter}) = 0.40 is close to unity for all the cases discussed. Note that the approximate equivalence of the elliptical shape and a randomly shaped cross-section for transmission of a circular mode is discussed in the Supplementary Methods.

\parskip = \baselineskip
\noindent
\textbf{Cross talk between axons}.
The neurons might be close to, or in contact with other neurons or non-neuronal cells in the brain (e.g. glia cells). Light in a myelinated axon would not leak out significantly, even if placed in direct contact with cells of lower refractive indices than the myelin sheath. However, if two or more myelinated axons are placed very close to each other (side by side), then light could leak out from one to the other. Supp.\ Fig.\ \ref{crosstalk}a shows the longitudinal refractive index profile of 2 axons (\textit{r}$'$ = \SI{4}{\micro\meter}) touching each other, and Supp.\ Fig.\ \ref{crosstalk}b is the EFPL (for those axons) when an input mode with wavelength \SI{0.4}{\micro\meter} is incident on one of them. In Supp.\ Fig.\ \ref{crosstalk}c, we notice that shorter wavelengths stay confined in the myelin sheath better, as expected. Supp.\ Fig.\ \ref{crosstalk}d--f deal with transmission (see Supplementary Methods for the procedure to quantify transmission) in the myelin sheath for different axon calibers, different wavelengths, and different separation between the axons. As a general rule, axons should be a wavelength away from one another to avoid cross talk, although the confinement for the same wavelength for different axon calibers can be quite different. Multimode waveguides (greater caliber) confine light much better than those with a few modes for a particular wavelength.

For our simulations, we considered cross talk between identical axons, which is stronger than that between non-identical ones. Also, the cross-talk between axons does not imply irretrievable loss. For perfectly identical optical fibers placed in contact, it is known that there is a complete power transfer from one to the other periodically \cite{snyder_S}. Moreover, extrapolation of the transmission for greater axon length is not straightforward, as light could propagate in the guided modes of the composite structure (many axons touching each other), with fluctuations (or oscillations) in power from one to the other. Since the most important source of loss (more so for the smaller wavelengths) here is light leaking into the myelin sheath of a different axon (and not the inside of the axons or outside), on average the power should be divided equally among the axons touching each other, provided that the segments in contact are long enough. Extrapolation from the data in Supp.\ Fig.\ref{crosstalk} as an exponentiation of the fraction of the power transmitted through \SI{50}{\micro\meter} should therefore be interpreted as a strict upper bound on the loss. Moreover, this might be a mechanism for information transfer between axons, leading to a collective behaviour of neurons in a nerve fiber (several axons bunched close together for a considerable length).

The power loss  when the axons touch each other under different spatial orientations is significantly less. For example, when two axons cross perpendicular to each other, the power loss is less than 0.5 $\%$ for all the axon calibers.

\begin{figure*}
\includegraphics[scale=0.35]{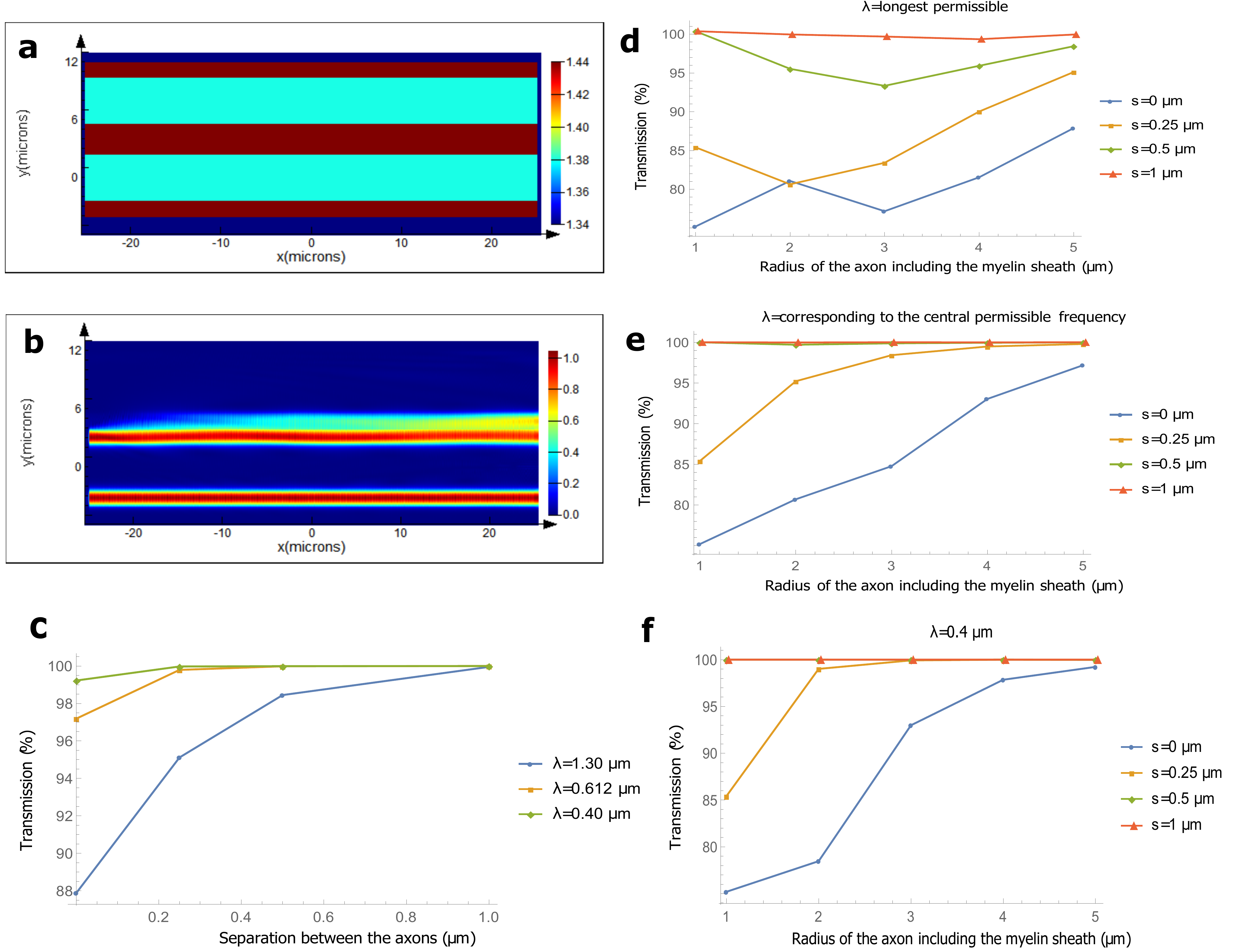}
\caption{\:\: \footnotesize{\textbf{Cross talk between axons.} (a) The refractive index profile of 2 axons touching each other (\textit{r}$'$ = \SI{4}{\micro\meter}). (b) EFPL as the input mode with wavelength \SI{0.4}{\micro\meter} crosses the region. (c) Transmission as a function of the separation between axons for different wavelengths (\textit{r}$'$ = \SI{5}{\micro\meter}). (d)-(f) Transmission as a function of the axon caliber for different wavelengths and different separation between axons.}}
\label{crosstalk}
\end{figure*}

\parskip = \baselineskip
\noindent
\textbf{Guided modes inside the axon}.
We have taken the refractive indices of the axon, the myelin sheath, and the medium outside as 1.38, 1.44, and 1.34 respectively for almost all our simulations. A vast majority of the modes of such a waveguide are confined strongly in the myelin sheath if it is thick enough. However, a few guided modes exist which have a greater fraction of optical power inside the axon than in the myelin sheath even if the myelin sheath is thick, and the wavelength is small. This is true if the axon has a greater refractive index than the medium outside the myelin sheath, and is sufficiently thick (true if the myelin is thick and the \textit{g-ratio} = 0.6). In the main text, we were particularly conservative and ignored the guided modes inside the axon, and treated them as loss, because we are not sure about the relevant light-guidance parameters inside the axon (see the later discussion on scatterers inside the axon). Without ignoring them, the transmission for all the inhomogeneities would be slightly better. Especially for the long paranodal regions, where some light inevitably leaks into the axon, one sees a clear difference.

\parskip = \baselineskip
\noindent
\textbf{Nodal and paranodal region with inclusion of the guided modes inside the axon}. Let's be optimistic and assume that the inside of the axon is homogeneous (has a constant refractive index of 1.38) to obtain an upper limit on the transmission as light crosses the nodal and paranodal regions. In Supp.\ Fig.\ \ref{paranodal_full}, we plot the modal transmission (power transmission in all the guided modes of the myelinated axon) after two paranodes and a node in between. We shall call two paranodes with a node in between a PNP (Paranode-Node-Paranode) region. We notice that for \textit{p-ratio} = 2.5, almost all the light for different axon calibers stays in the guided modes within a wavelength span from the myelin sheath (comparing it with Fig.\ 1 in the main text, where we took the transmission in the guided modes only upto a wavelength away from the myelin sheath boundaries). Also, for longer paranodal regions, the smaller wavelengths scatter more into the axon (and also in the medium outside the myelin sheath) than the longer wavelengths, as is evident from the difference in the transmission as compared to Fig.\ 2 in the main text. As an example, the transmission in all the guided modes for $\lambda$ = \SI{0.4}{\micro\meter} and \textit{r}$'$ = \SI{5}{\micro\meter} is 67.09 $\%$, but that within a wavelength span of the myelin sheath is only 33.78 $\%$. In a realistic scenario where there are scatterers inside the axon, the transmission would lie between these values. So, the plots in Supp.\ Fig.\ \ref{paranodal_full} should be interpreted as an upper bound on the transmission and Fig.\ 2 in the main text should be interpreted as a lower bound.

\begin{figure*}
\includegraphics[scale=0.18]{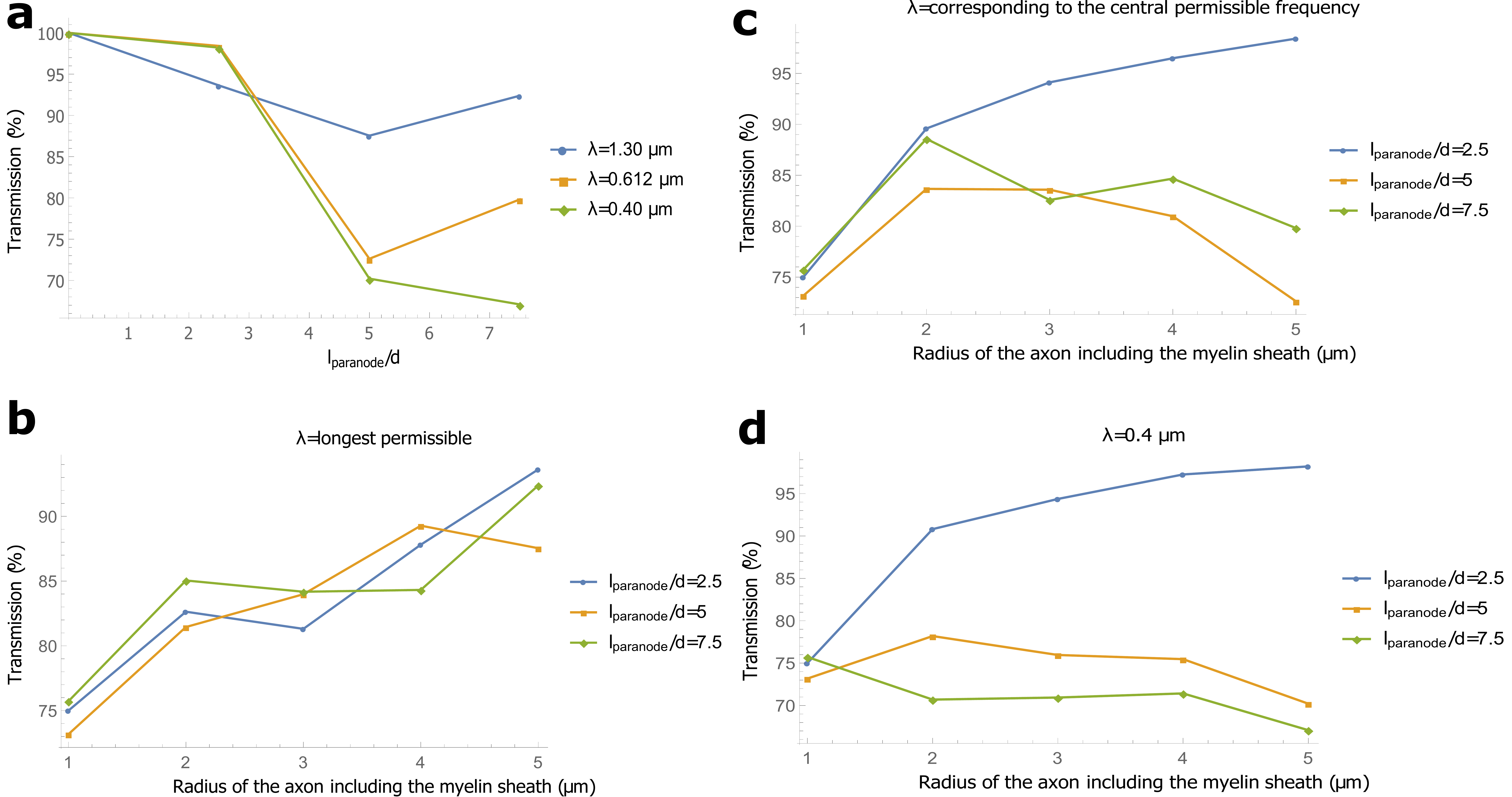}
\caption{\:\: \footnotesize{\textbf{Nodal and paranodal region with inclusion of the guided modes inside the axon.}  (a) Transmission in all the guided modes as a function of $l_{paranode}/d$ (\textit{p-ratio}), where $l_{paranode}$ is the length of a paranode, and \textit{d} is the thickness of the myelin sheath. The \textit{p-ratio} is varied by changing the length of the paranode, keeping the axon caliber constant ($r$ = \SI{3}{\micro\meter}, and $r'$ = \SI{5}{\micro\meter}). (b)-(d) Transmission in the guided modes as a function of the axon caliber for different wavelengths and different \textit{p-ratios}.}}
\label{paranodal_full}
\end{figure*}

\begin{figure}
\begin{center}
\includegraphics[scale=0.08]{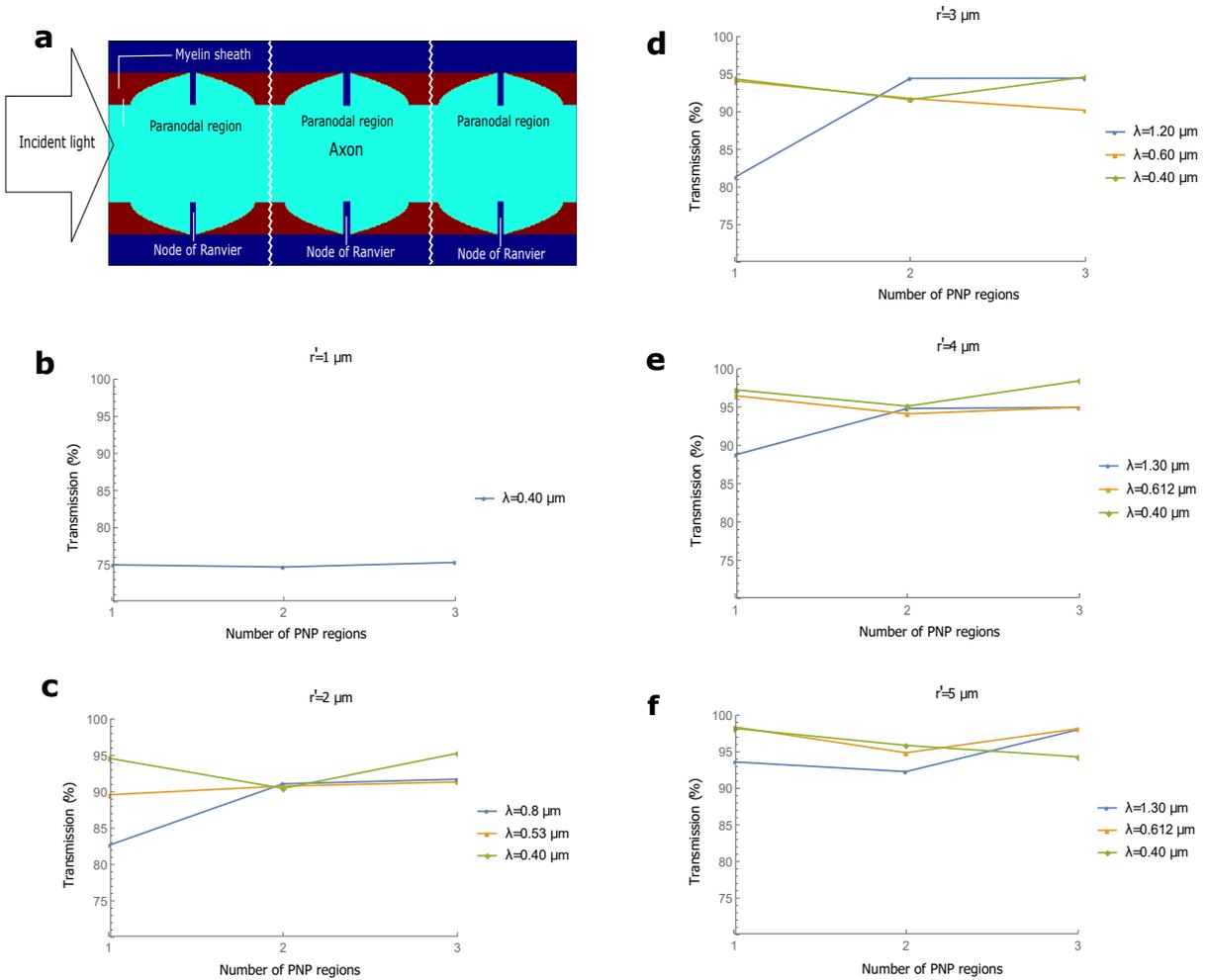}
\caption{\:\: \footnotesize{\textbf{Subsequent PNP regions.}  (a) Geometry of 3 PNP regions placed sequentially. A PNP (Paranode-Node-Paranode) region is defined as two facing paranodes with a node in between. The discontinuity between two PNP regions implies the presence of a straight and uniform internode there. (b)-(f) Transmission in all the guided modes as a function of the number of PNP regions for different wavelengths and different axon calibers. Note that the transmission is normalized to unity after each PNP region, such that the total modal transmission after 3 PNP regions is the product of the modal transmission after each of these regions. }}
\label{paranodal_iterations}
\end{center}
\end{figure}

\parskip = \baselineskip
\noindent
\textbf{Subsequent nodal and paranodal regions}. If the inhomogeneities in the rest of the internodal length is within the acceptable values, there would be no more loss as the rest of the light is in the guided modes. However, since there is mixing of modes as light passes through the paranodal regions, one might wonder how the mixture of modes behaves as it encounters the next PNP region (after an internodal length). Supp.\ Fig.\ \ref{paranodal_iterations} shows the transmission in the guided modes after subsequent PNP regions for different axon calibers and different wavelengths for \textit{p-ratio} = 2.5. Note that the transmission is re-normalized to unity after each PNP region, such that the total modal transmission after 3 PNP regions is the product of the modal transmission after each of these regions. In general, the longest permissible wavelengths (weakly confined) get better or almost saturate after 3 PNP regions. For shorter wavelengths, the modal transmission after each segment is less predictable since they are more prone to distortions in the shape of the myelin and undergo significant mode mixing. However, for most of the cases, the modal transmission fluctuates both ways (increases and decreases), and an average close to the first pass is approximately true. Thus, we can approximately predict the modal transmission after multiple PNP regions by exponentiating the modal transmission through one.

\noindent
\textbf{Effect of the scatterers and possibility of light guidance inside the axon}. There are many potential scatterers inside the axon, e.g. microtubules, mitochondria, agranular endoplasmic reticulum, and multivesicular bodies. We would not only need the refractive indices of these structures, but also their shapes, sizes and spatial distribution, to accurately predict their effect on light guidance. We have little relevant (and sometimes conflicting) data. For instance, Sato et al. measured the refractive index of microtubules to be 1.512 \cite{sato_S}, but Mershin et al. measured the refractive index of tubulin, the building block of microtubules to be 2.9 \cite{mershin_S}. Microtubules are one of the most numerous structures inside the axon, forming the cytoskeleton and a rail-road for the transport of materials inside the axon. The density of microtubules varies during the axon differentiation from $\sim$1 $\%$ in the initial phase to $\sim$3 $\%$ during the most dense phase and again drops (to a value we do not know) \cite{yu_S}.

To study the scattering effects of the microtubules on our previous simulations, we distribute them randomly (but according to a uniform distribution) such that they occupy $\sim$2 $\%$ of the volume inside the axon. Their refractive index is taken to be 1.5 and they are placed in a medium of refractive index 1.38. We had seen that in a few of our previous simulations, some fraction of optical power leaked into the axon, e.g. for large variation in the cross-sectional area, and paranodal regions. We ran the simulations again, this time in the presence of the microtubules. We found negligible variation in the transmission, both inside and outside the axon ($\pm \sim$1 $\%$). Even the light that leaked into the axon did not scatter much in the presence of the microtubules (owing to their small size and close to uniform distribution).

There are proposals of light guidance by the microtubules and mitochondria inside the axon \cite{thar_S,rahnama_S,jibu1_S,jibu2_S}. But they are too tiny for this to be realistic in the observed biophotonic wavelength range.  Mitochondria are typically less than a few microns long, and microtubules are too thin (tubular structures with the inner and outer diameters  as $\sim$12 nm and $\sim$24 nm respectively) to confine light in the biophotonic wavelength range (waveguide dimension should be comparable to the wavelength of light). However, if we assume that the microtubules are uniformly distributed, we can approximately average the refractive index of the composite system comprising of the axonal fluid and microtubules as $\sqrt{f\times n^2_m + (1-f) \times 1.34^2}$, where \textit{f} and $n_m$ are the volume fraction  and the refractive index of the microtubules respectively, and 1.34 is the refractive index of the fluid inside the axon. The average is possible since the microtubules are much smaller than the wavelength of light, and so is the average separation between them \cite{yu_S}. We could wonder whether this composite system can guide light, which is only possible if the inside of the axon has a higher refractive index than the medium outside. If the refractive index of the microtubules is 1.5, then a typical volume fraction, e.g. 1.7 $\%$ would give $n_{avg}$ = 1.343, and if the refractive index is 2.9, then $n_{avg}$ = 1.381.  Since the observed refractive indices inside the axon in both the longitudinal and the transverse directions are in a broad range (1.34--1.38 in \cite{antonov_S}, and 1.35--1.40 in \cite{wang_S}), assuming the axon as a uniform medium with refractive index 1.38 is not entirely correct. Moreover, the axons can be in direct contact with glia cells which can have comparable refractive indices as the inside of the axon. This would prevent guided modes to exist inside the axon. Note that if the refractive index of the axon is lower than 1.38, most of our simulations in the main text will yield slightly better transmission as the light guidance mainly depends on the refractive index contrast. And if the refractive index of the outside is greater than 1.34, the transmission will suffer slightly. However, since the refractive index of the myelin sheath is much larger than both the regions, these effects would not be too significant for most of the simulations.

However, if we assume that the axons are not in contact with other glia cells, and have a higher refractive index than  the interstitial fluid outside, then weak guidance might still be possible if the mode does not scatter off of the bigger (but less numerous) scatterers (e.g. mitochondria, Endoplasmic Reticulum, and vesicles). We do not know the volume fraction of these scatterers precisely but some work, e.g.\ \cite{zelena_S} suggest that they occupy at least 10 $\%$ of the volume. We model these scatterers as ellipsoids with the 3 semi-minor axes ranging from \SI{0.1}{\micro\meter} to \SI{0.4}{\micro\meter}, \SI{0.1}{\micro\meter} to \SI{0.4}{\micro\meter}, and \SI{1}{\micro\meter} to \SI{3}{\micro\meter} respectively and place them in axon with \textit{r}$'$ =  \SI{5}{\micro\meter}. Their refractive indices are taken to be 1.4. Let's take 2 different values of the refractive index of the axon. For a value 1.38, the total power transmission (calculated by integrating the real part of the Poynting vector of the output light directly across the required area, and dividing it by the source power) upto a wavelength away from the axonal boundaries in a \SI{100}{\micro\meter} long structure for a mode  confined inside the axon at wavelength \SI{0.612}{\micro\meter} is 75.47 \%, while for \SI{1.3}{\micro\meter} wavelength, the transmission is 95.93 \%. If the axon's refractive index is 1.35, then the transmission for the wavelength \SI{0.612}{\micro\meter} is 16.06 \%, while no guided modes exist for the wavelength \SI{1.3}{\micro\meter}. A lower density of these scatterers, or smaller sizes, (or larger wavelengths than \SI{0.612}{\micro\meter}) would, of course yield greater transmission. The transmission for \SI{0.612}{\micro\meter} wavelength light is different for different refractive index values of the axon because scattering depends strongly on the refractive index contrast. A mitochondrion (refractive index 1.40) placed in a medium with refractive index 1.35 would act as a much stronger scatterer than if placed in a medium with refractive 1.38. Thus, an average uniform refractive index of 1.38 for the axon might still guide light at large wavelengths, but an average uniform refractive index of 1.35 seems more believable (assuming the refractive index of microtubules to be $\sim$1.50). In this case, the bigger scatterers lead to significant loss, even if the microtubules themselves do not. Therefore, we do not believe that there could be guided modes inside the axon which can transmit efficiently.

We again ran many of our previous simulations (with the input mode confined primarily in the myelin sheath) in the presence of all these scatterers inside the axon. We varied the refractive index of the axon from 1.34 (the refractive index of the medium outside) to 1.38. We verified that light well confined in the myelin sheath does not see these scatterers at all. Even when the light diverges into the axon because of the geometry of the structure (e.g. the varying cross-sectional area), there is still not a dramatic variation in the transmission. Both the transmission in the myelin sheath up to a wavelength away from the boundaries, and the total transmission across the whole cross-section (including the inside of the axon) do not change greatly; the observed variation was on the order of a few percent. Note that for a few simulations, $\sim$15--20 $\%$ of the fraction of output light can be inside the axon. The light diverging inside need not even be in the guided modes of the waveguide. This runs counter to intuition, since we saw that a guided mode inside the axon scattered badly. This unintuitive phenomenon can be explained again by the unconventional nature of this waveguide, where all the light leaking inside is not irretrievably lost (even if it is not in the guided modes of the structure). It can come back to the myelin sheath without interacting strongly with the scatterers. This shows that we might have been too conservative while considering the power only within a wavelength of the myelin sheath boundaries. However, there still might be other phenomena happening (e.g. absorption) inside the axon, and we prefer to be cautious about the inside.

Next, we shall see how varying the refractive index of the axon affects the transmission of a mode (confined primarily in the myelin sheath) in the PNP region.

\begin{figure*}
\includegraphics[scale=0.26]{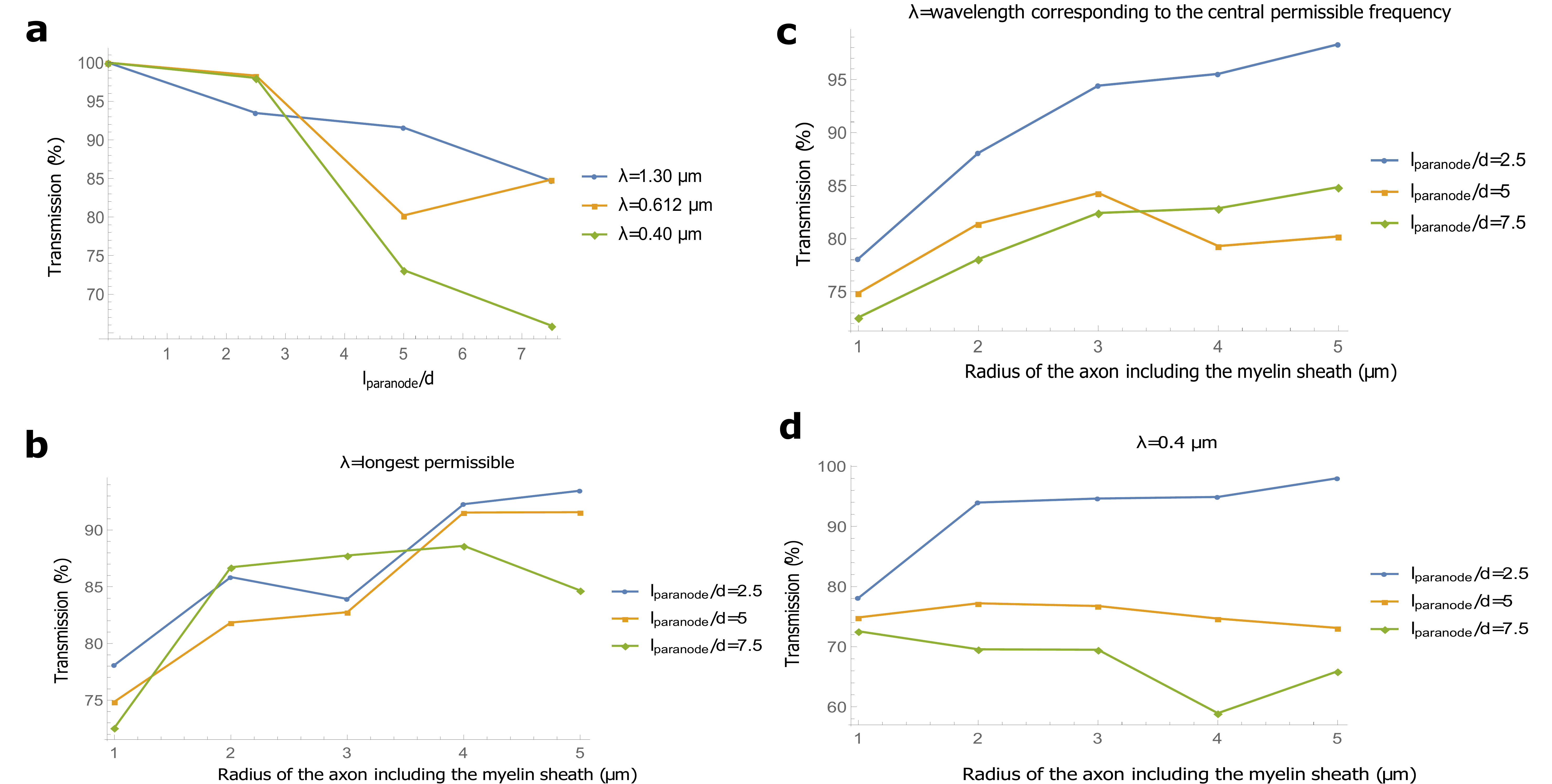}
\caption{\:\: \footnotesize{\textbf{Nodal and paranodal regions for a different set of refractive indices.} The refractive indices of the axon and the cytoplasmic loops are taken as 1.34 and 1.38 respectively. (a) Transmission in all the guided modes as a function of the \textit{p-ratio} for different wavelengths (\textit{r}$'$ = \SI{5}{\micro\meter}). (b)-(d) Transmission in the guided modes as a function of the axon caliber for different wavelengths and different paranodal lengths.}}
\label{paranodal_dri2}
\end{figure*}

\noindent
\textbf{Varying the refractive indices of the axon and the cytoplasmic loops.} We have observed that the paranodal regions might be the main contributor to loss (if the other inhomogeneities are low). For our simulations so far, we have assumed that the refractive index of the cytoplasmic loops is the same as that of the axon (1.38). As far as we know, no direct measurement of the refractive index of these loops has been performed, but they are however considered \textit{`dense'} \cite{chang_S}.Since these loops are part of glia cells, which usually have higher refractive indices, these loops might have higher refractive indices than the inside of the axon too.  In Supp.\ Fig.\ \ref{paranodal_dri2}, we show the result of another set of simulations where the refractive index of the axon is kept the same as the medium outside (1.34), and that of the loops is higher (1.38). We find that in almost all the cases (different paranodal lengths, different wavelengths, and different axon calibers), the transmission in the guided modes is higher as compared to the previous set of simulations (see Fig.\ 1 in the main text). If cytoplasmic loops have a higher refractive index, then they prevent the mode from diverging into the axon, and serve as weak waveguides themselves. Note that for the same refractive index of the axon and the cytoplasmic loops (e.g. 1.35 each), the results would be similar to those when both had their refractive indices 1.38.

\noindent
\textbf{Mathematics of mode expansion and transmission calculation.}
We have often mentioned the expansion of the output field in the basis of guided modes, and the calculation of the transmission by evaluating the Poynting vector, integrating its real part across the area of interest and dividing it by the input power. Here we give the mathematics of these procedures.

Let the electric (E) and magnetic (M) field profiles (frequency domain) of the light incident in the axon be denoted by $\vv{E}_{in}$, and $\vv{H}_{in}$ respectively, and the field profiles of the light at the terminal end of the axon segment in the transverse plane (perpendicular to the length) be denoted by $\vv{E}_{out}$, and $\vv{H}_{out}$ respectively. We can express
\begin{equation}
\begin{split}
\vv{E}_{out}= \vv{E}_{guided} + \vv{E}_{non-guided}\\
\vv{H}_{out}= \vv{H}_{guided} + \vv{H}_{non-guided}
\end{split}
\end{equation}
where $\vv{E}_{guided}$, and $\vv{H}_{guided}$ are the fields for the fraction of light in the finite number of guided modes of the waveguide, and $\vv{E}_{non-guided}$, and $\vv{H}_{non-guided}$ are the fields for the fraction in the infinite number of non-guided modes. Light in the non-guided modes of a uniform structure would be lost eventually. The guided part can further be expanded as
\begin{equation}
\begin{split}
\vv{E}_{guided}= \sum_i (a_i \vv{E}^{forward}_{i} + b_i  \vv{E}^{backward}_{i})\\
\vv{H}_{guided}= \sum_i (a_i \vv{H}^{forward}_{i} - b_i  \vv{H}^{backward}_{i})\\
\end{split}
\end{equation}
where $\vv{E}_{i}$, and $\vv{H}_{i}$ are the fields corresponding to a guided mode $\phi_i$, and $a_i$ and $b_i$ are the transmission coefficients for the forward and backward propagating waves respectively. The summation is over the entire set of the orthogonal guided modes of the structure. The coefficients are given in terms of the overlap integrals as
\begin{equation}
\begin{split}
a_i=0.25 \times (\frac{\int (\vv{E}_{guided} \times \vv{H}^*_{i} )\cdot \vv{dS}}{P_i}+\frac{\int (\vv{E}^*_{guided} \times \vv{H}_{i} )\cdot \vv{dS}}{P^*_i})\\
b_i=0.25 \times (\frac{\int (\vv{E}_{guided} \times \vv{H}^*_{i} )\cdot \vv{dS}}{P_i}-\frac{\int (\vv{E}^*_{guided} \times \vv{H}_{i} )\cdot \vv{dS}}{P^*_i})
\end{split}
\end{equation}
where $\mathrm{\vv{dS}}$ is the differential area element in the transverse plane of interest, and the complex power of the $\mathrm{i^{th}}$ mode $\phi_i$ is
\begin{equation}
P_i=0.5\times \int (\vv{E}_{i} \times \vv{H}^*_{i})\cdot \vv{dS}
\end{equation}

The percentage transmission into all the guided modes of the structure is given by
\begin{equation}
T=\frac{0.5\times \int Re(\vv{E}_{guided} \times \vv{H}^*_{guided})\cdot \vv{dS}}{0.5\times \int Re(\vv{E}_{in} \times \vv{H}^*_{in})\cdot \vv{dS}} \times 100
\end{equation}
Here, $\vv{S}_{guided}=\vv{E}_{guided} \times \vv{H}^*_{guided}$ is the  time averaged Poynting vector for the guided fraction of the output light, and $Re()$ denotes the real part. Integration of the real part of the Poynting vector across an area quantifies the time-averaged power flow through that area, while the integration of the imaginary part quantifies the reactive power (e.g. because of interference due to a standing wave).

In specific contexts (in particular after the PNP regions, see Fig.\ 2 in the main text), we integrate the real part of the Poynting vector (with the electromagnetic fields corresponding to the guided portion of the output light) across the myelin sheath up to a wavelength away from the boundaries to obtain the percentage transmission
\begin{equation}
T=\frac{0.5\times \int_{\rho=r-\lambda}^{\rho=r+\lambda} Re(\vv{E}_{guided} \times \vv{H}^*_{guided})\cdot \vv{dS}}{0.5\times \int Re(\vv{E}_{in} \times \vv{H}^*_{in})\cdot \vv{dS}} \times 100
\end{equation}
where $\rho$ is the radial coordinate, $\lambda$ is the wavelength, and $r$ and $r'$ are the inner and outer radius of the myelin sheath as defined earlier. We include only the guided fraction of the light because the non-guided fraction is expected to decay over the course of the long internode following the PNP region (provided that the internode is approximately uniform).

In certain other instances (e.g. varying cross-sectional area and shape), where the cross-section continuously changes, some fraction of light in the non-guided modes at a particular cross-section might be included in the the basis of guided modes at an adjoining cross-section and vice-versa. Therefore, it is more appropriate to observe the total power transmission (up to a wavelength of the myelin sheath boundaries) instead of the modal transmission.  In such cases we integrate the real part of the Poynting vector with the fields corresponding to the output light directly to obtain the percentage transmission
\begin{equation}
T=\frac{0.5\times \int_{\rho=r-\lambda}^{\rho=r+\lambda} Re(\vv{E}_{out} \times \vv{H}^*_{out})\cdot \vv{dS}}{0.5\times \int Re(\vv{E}_{in} \times \vv{H}^*_{in})\cdot \vv{dS}} \times 100
\end{equation}
\noindent
\textbf{\large{Supplementary Methods}}
\parskip = \baselineskip

\noindent
\footnotesize{\textbf{Continuously varying non-circular cross-sectional shape}.
We simulate the effect of the change in the cross-sectional shape of an axon in the longitudinal direction by twisting an elliptical axon. The semi-major and the semi-minor axes of the ellipse (\textit{a} and \textit{b} resp.) at x = 0 (the starting point of the axon) are changed for different simulations. We incident an eigenmode of a circular axon with $r = (a + b)/2$, and $r/r'=0.6$. The myelin sheath boundary is another ellipse with its axes, $a' = a+d$ and $b' = b+d$, where $d = r'-r$. The myelin sheath is thus an approximate parallel curve to the axon. The shape of the axon changes continuously such that at one-fourth of the axonal segment, it becomes a perfect circle with radius $r = (a + b)/2$, at half the length, it interchanges its axes, and at the end of the segment (\SI{50}{\micro\meter}), it resumes its original shape. The area of the cross-section remains almost constant by this twist (less than 10 $\%$ variation for all the simulations). Different values of the change in the aspect ratio ($\Delta AR$) are obtained by adopting the same procedure for ellipses with different semi-axes.

An approximate equivalence between an elliptical shape and a random cross-sectional shape (as in the main text) can be established. The equation of an ellipse in polar coordinates is $ \rho(\theta) = a b/(\sqrt{(b \cos{\theta})^2+(a \sin{\theta})^2})$, where $\rho$ is the radial coordinate and $\theta$ is the polar angle from the $a$ axis. The mean of the distance of the points from the origin is very close to $r = (a + b)/2$ (less than 7 $\%$ variation for all the simulations). In the main text, we generated random points according to a Gaussian distribution along the circumference of the cross-section, and the s.d. of the separation of those points from a circle of radius \textit{r} is taken as the degree of inhomogeneity. For an ellipse, the s.d. of separation from a circle of radius $r = (a + b)/2$ can similarly be calculated as $\sqrt{1/(2 \pi) \int_{0}^{2 \pi} ((a+b)/2-a b/(\sqrt{(b \cos{\theta})^2+(a \sin{\theta})^2}))^2 d\theta}$ . We compare transmission in an elliptic (non changing cross-sectional area) waveguide, and a waveguide with an arbitrary cross-sectional area with the same s.d for some of the simulations, and find that there is comparable or higher loss in an elliptical waveguide. This suggests that an axon with changing cross sectional shape (random) along its length might also undergo similar loss as a twisting elliptical axon. We quantify the change in aspect ratio ($\Delta AR$) as a measure of the change in the cross-sectional shape for elliptical shapes. For example, if the cross-section is an ellipse with \textit{a} = \SI{3.9}{\micro\meter}, and \textit{b} = \SI{2.1}{\micro\meter} at x = 0, after the twisting procedure, $\Delta AR = 2 \times (3.9/2.1-2.1/3.9)= 2.64$ (the factor 2 shows that it is twisted to get back to the original shape after the segment).

The transmission is calculated by integrating the real part of the Poynting vector across an area between 2 ellipses, one with the semi-axes $a+d+\lambda$, $b+d+\lambda$, and the other with the semi-axes, $a-\lambda$, and $b-\lambda$, where $\lambda$ is the wavelength of the light, and the other symbols hold their previous meanings. The procedure adopted to account for the change in the mode profiles with wavelength is the same as discussed in the Methods of the main text (e.g. as in bends). We divide the transmission for the larger wavelengths by the transmission within a wavelength of the myelin sheath for a circular waveguide on sending in a mode with the central permissible frequency to obtain the normalised transmission.  The losses are in fact a combination of the insertion loss (coupling loss of the input light to the first cross section it sees) and the propagation loss (can be understood as coupling losses for subsequent cross-sections), but as a conservative approach, we allocate everything to the propagation loss. Under this assumption, we expect that an ellipse with a larger (or smaller) aspect ratio ($a/b$) to start with, would have almost similar transmission if $\Delta AR$ is the same (for the same mean caliber \textit{r}, i.e. (\textit{a}+\textit{b})/2).  For a waveguide with arbitrary cross-sectional shape that changes continuously, an analogous picture (to the twisting of an elliptical waveguide) is to start with some random shape, then reduce the randomness to reach a perfect circular shape, then increase the randomness again to arrive at a shape with the axes reversed (a $\pi/2$ rotated form of the original shape), and carry out this procedure again to arrive at the original shape at the end of \SI{50}{\micro\meter}.

\parskip = \baselineskip
\noindent
\footnotesize{\textbf{Cross-talk between axons}. We  place two identical axons side by side, send in light through one of them and note the power (by integrating the real part of the Poynting vector across the myelin sheath only) transmitted across the same axon in which the mode was incident. We divide the power for each wavelength by the power transmitted in the myelin sheath alone (not up to a wavelength) in the absence of the second axon to obtain the normalised transmission.}

\end{document}